\pdfoutput=1
\documentclass[aps,prd,amsmath,floats,floatfix, twocolumn,
superscriptaddress,nofootinbib,showpacs,longbibliography]{revtex4-1}
\usepackage[T1]{fontenc}
\usepackage[utf8]{inputenc}
\DeclareUnicodeCharacter{2009}{\nobreakspace}
\usepackage{lmodern}
\usepackage{times}
\usepackage[varg]{txfonts}
\usepackage[normalem]{ulem}
\usepackage{verbatim}
\usepackage[dvipsnames, usenames]{xcolor}
\definecolor{linkcolor}{rgb}{0.0,0.2,0.5}
\usepackage[hypertexnames=false, unicode, colorlinks=true, linkcolor=linkcolor,
citecolor=linkcolor, filecolor=linkcolor,urlcolor=linkcolor,
pdfusetitle]{hyperref}
\usepackage[all]{hypcap}
\usepackage{graphicx}
\usepackage{xspace}
\usepackage{amssymb}
\usepackage{bm} 
\usepackage{microtype}
\usepackage[english]{babel}
\usepackage{aas_macros}
\usepackage{physics}
\usepackage{orcidlink}
\usepackage{multirow}
\usepackage{enumitem}
\DeclareMathAlphabet{\mathpzc}{OT1}{pzc}{m}{it}

\begin{document}

\preprint{APS/123-QED}

\title{Periodic line-of-sight velocity-driven modulations to gravitational waves emitted by compact binaries in Keplerian outer orbits}

\author{Avinash Tiwari\,\orcidlink{0000-0001-7197-8899}}
\email{avinash.tiwari@iucaa.in}
\affiliation{Inter-University Centre for Astronomy and Astrophysics, Post Bag 4, Ganeshkhind, Pune - 411007, India}

\author{Shasvath J. Kapadia\,\orcidlink{0000-0001-5318-1253}}
\email{shasvath.kapadia@iucaa.in}
\affiliation{Inter-University Centre for Astronomy and Astrophysics, Post Bag 4, Ganeshkhind, Pune - 411007, India}

\author{Aditya Vijaykumar\,\orcidlink{0000-0002-4103-0666}}
\email{avijaykumar@cita.utoronto.ca}
\affiliation{Canadian Institute for Theoretical Astrophysics, University of Toronto, 60 St. George Street, Toronto, ON M5S 3H8, Canada}

\author{Sourav Chatterjee\,\orcidlink{0000-0002-3680-2684}}
\email{sourav.c@tifr.res.in}
\affiliation{Tata Institute of Fundamental Research, Homi Bhabha Road, Navy Nagar, Colaba, Mumbai 400005, India}

\hypersetup{pdfauthor={Tiwari et al.}}

\begin{abstract}
The centre of mass (CoM) of compact binary coalescences (CBCs) occurring in the vicinity of a supermassive black hole, through interaction with an arbitrary third body (e.g., of stellar mass), or in a dense stellar environment, will undergo a time-varying line-of-sight (LOS) velocity. This in turn leads to a time-varying Doppler shift and corresponding modulations in the shape of the gravitational waves (GWs). The phase and amplitude corrections arising from constant LOS acceleration and its higher-order time derivatives are already known. Specifically, these effects lead to corrections to the GW waveform at $-4n$ post-Newtonian (PN) order, where $n$ is the $n^{th}$ time derivative of the LOS velocity. In the context of a circular or eccentric outer orbit of the CoM of the CBC, these effects can be thought of as approximations to the LOS velocity in the limit: observation duration $\ll$ period of the outer orbit. However, this condition is not necessarily always satisfied. In this {\it paper}, we present phase and amplitude corrections to the GW waveforms arising from a periodic non-relativistic LOS velocity for circular and eccentric outer orbits of the CBC's CoM. Specifically, these lead to phase and amplitude modulations at 4 PN order, and reduce to the known corrections for constant kinematic parameters under appropriate limits mentioned above. We also perform a Fisher matrix analysis to forecast constraints on the environment that is sourcing the time-varying LOS velocity, for various future ground and space-based detectors. We further show that constraints acquired using GW waveforms derived in this work improve significantly in comparison to those acquired from approximate methods valid for constant kinematic parameters.
\end{abstract}

\maketitle

\section{Introduction}\label{sec:intro}
The LIGO-Virgo-KAGRA (LVK) gravitational-wave (GW) detector network~\cite{LIGODetector, VirgoDetector, KAGRADetector, PhysRevD.88.043007} has detected over 300 compact binary coalescence (CBC) events~\cite{LIGOScientific:2026sit, LIGOScientific:2025slb}.
The provenance of these CBCs remains an active area of investigation (see, e.g., Ref.~\cite{Mapelli:2021taw} for a review), with numerous studies dedicated to identifying their formation channels.  In the absence of electromagnetic (EM) counterparts~\cite{PhysRevLett.119.161101, Abbott_2017:GW170817, KAGRA:2021vkt}, probing the environment of a GW merger becomes very challenging. This problem becomes even more severe because of the poor sky localization of the GW events~\cite{Chen:2016tys}. Information about putative formation channels on a population level~\cite{LIGOScientific:2025pvj, LIGOScientific:2026ctl} can still be speculated based on the intrinsic properties of the objects~\cite{Zevin:2020gbd, Pierra:2024fbl}, such as mass ratios, eccentricities, and spins. But one cannot, in general, pinpoint a formation channel on a single event basis. 

In principle, single-event host identification could be achieved by studying the kinematics of the centre-of-mass (CoM) of CBCs. Previous studies~\cite{Yunes2011, Bonvin2017, Vijaykumar_2023, PhysRevD.110.083008, He:2026bzv, Zhao:2026ydo, Roy:2026mco, Pompili:2026tdf, Roy:2026duh, Gera:2025ugl} focused on their line-of-sight velocity (LOSV) varying linearly with time, i.e, modulations incurred due to a constant line-of-sight acceleration (LOSA) (see Refs.~\cite{Vijaykumar_2023, 2024arXiv240101743H, Hendriks:2026kys, Roy:2026mco, Pompili:2026tdf, Roy:2026duh, Pathak:2026cik, LIGOScientific:2026qni, LIGOScientific:2026fcf} for constraints on LOSA from GW events). In fact, Refs.~\cite{Yunes2011, Bonvin2017, Vijaykumar_2023, PhysRevD.110.083008} and~\cite{Tiwari_pipe_2026} showed that the LOSA of the CoM of a CBC leads to modulation in the GW waveform at $-4$ post-Newtonian (PN) order, Refs.~\cite{Santos:2025ass, Hendriks:2026kys, Tagawa:2025tfd, Takatsy:2025bfk, Hendriks:2024gpp, Giri:2026wgy, Camilloni:2023xvf, Cocco:2025adu, Cocco:2025udb} investigated the effects of the environment, including LOSA, on the CBC, while Ref.~\cite{Tiwari:2023cpa} explored the prospects for detecting the LOSA of a CBC's CoM in globular clusters using space-based detectors. Moreover, Refs.~\cite{Tiwari:2024pvb} and~\cite{Tiwari:2025qqx} showed that by studying the kinematics of the CoM of the CBC through the imprints of LOSA and other higher-order time derivatives of the LOSV onto the GW waveform,  one can profile the merger environments and hence determine the CBC's formation channel on a single event basis.

In Ref.~\cite{Tiwari:2024pvb}, we assumed that outer orbital periods of the CBC's CoM in circular and eccentric outer orbits --- henceforth COO and EOO, respectively --- are $\gg$ the observation time, or equivalently, that we observe only some segment of the outer orbit. This allowed us to Taylor-expand the LOSV of the CoM of the CBC in terms of LOSA and its other higher-order time derivatives. However, the method is valid only when the CBC is in a very wide outer orbit, i.e., it is far away from an SMBH/third body. This forbids analysis of a significant portion of parameter space closer to the third body.

In this paper, we extend the formalism to periodic non-relativistic LOSVs for circular and eccentric outer orbits. We calculate the frequency-domain corrections to the GW phase and amplitude as a function of the LOSV parameters and show that we can extract the information about the mass of the third body in the vicinity of a CBC, the radius/semi-major axis of the outer orbit, and the eccentricity of the outer orbit. We consider various single-detector configurations, viz., a LIGO detector in O5 at A+ sensitivity \cite{aplus}, the Einstein Telescope (ET) \cite{punturo2010} of the XG network \cite{Reitze:2019iox, punturo2010}, the LISA \cite{LISA:2017pwj} and DECIGO \cite{Sato:2017dkf} space-based detectors.

\section{Phase and Amplitude Corrections}\label{sec: ph_amp_corr}
\begin{figure}[!ht]
    \centering
    \includegraphics[width=0.995\linewidth]{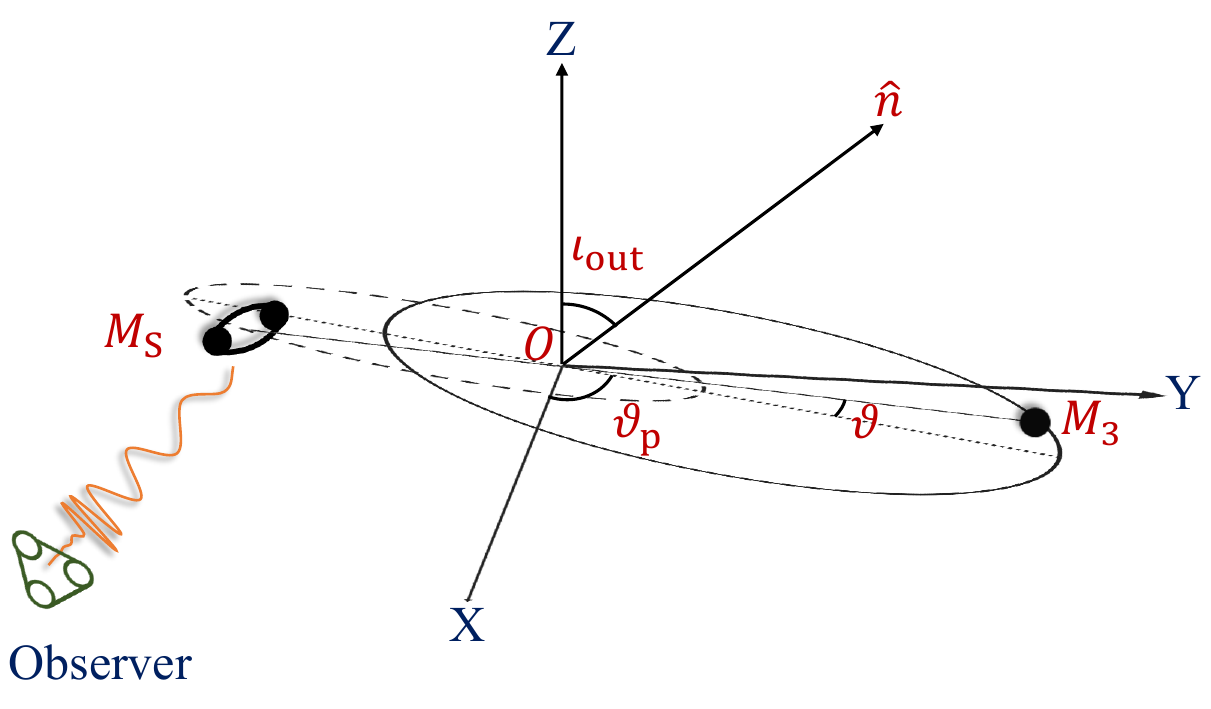}
    \caption{The schematic representation of a BH ($M_3$) and a BBH ($M_{\rm S}$) orbiting in eccentric orbits around the system's centre of mass (barycenter) $O$. $\vartheta_{\rm p}$ is the longitude of periapsis, $\vartheta$ is the true anomaly of the outer orbit, and $\iota_{\rm out}$ is the angle between the angular momentum (along the Z-axis) of the outer orbit and the observer's LOS $\Hat{n}$.}
    \label{fig: schem_bbh_bh}
\end{figure}

Let $M_3$ be the mass of a third body in the vicinity of a CBC of component masses\footnote{All three masses are in the source frame.} $m_{1,\rm S}$ and $m_{2,\rm S}$ and the CoM of the CBC be in an outer orbit around the system's barycentre (see Figure~\ref{fig: schem_bbh_bh}), $V_{\rm LE}$ and $V_{\rm LC}$ be the LOSVs of its CoM when the orbit is eccentric and circular, respectively, and $e_{\rm out}$ be the eccentricity of the outer orbit. Then we can write~\cite{Exoplanets_s_seager_book_2010}
\begin{align}
    \label{eq: losv_EO}
    V_{\rm LE} &= \frac{V_{\rm L,0}  \left[\cos \left(\vartheta+\vartheta_{\rm p}\right) + e_{\rm out} \cos \vartheta_{\rm p}\right]}{\sqrt{1-e_{\rm out}^2}} \\
    V_{\rm LC} &= V_{\rm L,0} \cos (\Omega_{\rm det} (t_{\rm u} - t_{\rm c}) + \theta_{\rm c})
\end{align}
where $V_{\rm L,0} / \sqrt{1 - e_{\rm out}^2}$ and  $V_{\rm L,0}$ are the maximum LOSVs in the eccentric and circular outer orbits, respectively, $\vartheta_{\rm p}$ is the longitude of periapsis\footnote{In the most general case, it would be the argument of periapsis. Since we do not consider the precession of periapsis or the change in the longitude of the ascending node $\vartheta_{\rm asc}$, $\vartheta_{\rm asc}$ will be a constant. For simplicity, we have set $\vartheta_{\rm asc} = 0$.} (see Figure~\ref{fig: schem_bbh_bh}), $\vartheta$ is the true anomaly, $\Omega_{\rm det} \equiv \Omega / (1 + z_{\rm cos})$\footnote{Note that the extra factor $1 / (1 + z_{\rm cos})$ is due to cosmological time dilation.} with $\Omega$ being the mean motion (angular frequency) of the eccentric (circular) outer orbit, $t_{\rm u}$ the observer-frame time accounting only for cosmological redshift, $t_{\rm c}$ the coalescence time, and $z_{\rm cos}$ is the cosmological redshift. 

Let $M = m_1 + m_2$ be the cosmologically redshifted total mass of the CBC, where $m_1 = m_{1,\rm S} (1 + z_{\rm cos})$ and $m_2 = m_{2,\rm S} (1 + z_{\rm cos})$, $M_{\rm LC}$ and $M_{\rm LE}$ be the detector frame total masses of the same after incorporating the Doppler shifts due to the LOSV of its CoM in case of the COO and EOO, respectively, and $z_{\rm LC} \equiv  V_{\rm LC} / c$ and $z_{\rm LE} \equiv  V_{\rm LE} / c$ be the corresponding Doppler shifts. Let $z_{\rm L,0} \equiv V_{\rm L,0} / c$ be the maximum Doppler shift in the case of the COO, where $c$ is the speed of light. Under the assumption that $z_{\rm L, 0} \ll 1$\footnote{In this analysis, we choose this number to be 0.05, as also done in Ref.~\cite{Tiwari:2024pvb}.} for COOs and $z_{\rm L, 0}/\sqrt{1 - e_{\rm  out}^2} \ll 1$ for EOOs, we can write
\begin{align}
    \label{eq: redshifted_mass}
    M_{\rm LE} &= M(1 + z_{\rm LE}), \\
    M_{\rm LC} &= M(1 + z_{\rm LC}).
\end{align}
These time-varying redshifts break the mass-redshift degeneracy and lead to modulations in the GW waveform. Under the stationary phase approximation (SPA), the GW waveform of the CBC moving with a time-varying relative LOSV can be written as
\begin{equation}
    \label{eq: WF}
    \tilde{h}_{\rm TV}(f) = \tilde{h}(f) \left( 1 + \frac{\Delta \mathcal{A}}{\mathcal{A}} \right)  e^{i \Delta \Psi (f)}
\end{equation}
where $\tilde{h}(f)$ is the unmodulated GW waveform~\cite{PhysRevD.80.084043}, $\mathcal{A}$ being the amplitude of the same, $f$ is the GW frequency, and $\Delta \Psi (f)$ and $\Delta \mathcal{A}/\mathcal{A}$ are the phase and amplitude corrections, respectively, due to time-varying LOSV. Following~\cite{Vijaykumar_2023, Tiwari_pipe_2026}, we derive these corrections in the upcoming sections. Specifically, these corrections appear at 4 PN order. Note that we derive all the corrections for the $(\ell,m) = (2,2)$ mode. However, the phase corrections can be trivially extended to higher modes by using the transformation $\Delta \Psi (f) \to (m/2) \Delta \Psi (2f/m)$ for the $m^{th}$ mode. However, a straightforward transformation does not exist for amplitude corrections, which would therefore require a mode-by-mode computation.

\subsection{Circular Outer Orbits}\label{subsec: circ_der}
For CBCs in circular outer orbits, we can write the Doppler shift due to the LOSV of the CBC's CoM defined above as:
\begin{equation}
    \label{eq: red_circ}
    z_{\rm LC} = z_{\rm L,0} \cos (\Omega_{\rm det} (t_{\rm u} - t_{\rm c}) + \theta_{\rm c}).
\end{equation}
Let $f_{\rm u}$ and $f_{\rm o}$ be the unperturbed and perturbed GW frequencies, respectively. Then Equations~A1-A3 of \cite{Vijaykumar_2023} (see also \cite{Tiwari_pipe_2026}) take the form:
\begin{align}
    \label{eq: dop_shift_circ}
    f_{\rm u} &= f_{\rm o} \left(1 + z_{\rm LC} \right) \\
    dt_{\rm u} &= \frac{dt_{\rm o}}{\left(1 + z_{\rm LC} \right)} \\
    v_{\rm u} &= v_{\rm o} \left(1 + z_{\rm LC} \right)^{1/3}
\end{align}
while Equation~A4 of the same takes the form:
\begin{multline}
    \label{eq: dvdt_circ1}
    \frac{dv_{\rm o}}{dt_{\rm o}} = \frac{1}{3}v_{\rm o} z_{\rm L,0} \Omega_{\rm det}  \sin (\Omega_{\rm det} (t_{\rm u} - t_{\rm c}) + \theta_{\rm c}) \\ + \left(1 + z_{\rm LC} \right)^{-4/3} \frac{dv_{\rm u}}{dt_{\rm u}}
\end{multline}
where $v_{\rm u} \equiv (\pi G M f_{\rm u} / c^3)^{1/3}$ and $v_{\rm o} \equiv (\pi G M f_{\rm o} / c^3)^{1/3}$, $G$ being the Gravitational constant. $dv_{\rm u}/dt_{\rm u}$ is given by Equation 3.6 of~\cite{PhysRevD.80.084043} with $v$ replaced by $v_{\rm u}$ and $\nu$ by the symmetric mass ratio $\eta$ defined as $m_1 m_2 / M^2$. Substituting the leading order term in Equation~\eqref{eq: dvdt_circ1} together with $v_u$ from Equation~\eqref{eq: dop_shift_circ}, we can write
\begin{multline}
    \label{eq: dvdt_circ2}
    \frac{dv_{\rm o}}{dt_{\rm o}} = \frac{1}{3}v_{\rm o} z_{\rm L,0} \Omega_{\rm det}  \sin (\Omega_{\rm det} (t_{\rm u} - t_{\rm c}) + \theta_{\rm c}) \\ + \left(1 + \frac{5}{3}z_{\rm L,0} \cos (\Omega_{\rm det} (t_{\rm u} - t_{\rm c}) + \theta_{\rm c}) \right) \frac{32 \eta}{5} \frac{c^3}{G M} v_{\rm o}^9
\end{multline}
As an first order of approximation for $t_{\rm u} - t_{\rm c}$, we use Equation 3.8b of~\cite{PhysRevD.80.084043} with $v$ and $\nu$ replaced by $v_{\rm o}$ and $\eta$, respectively. We then invert Equation~\eqref{eq: dvdt_circ2} to obtain $dt_{\rm o} / dv_{\rm o}$ and integrate it in the limits $v_o \to v \equiv (\pi G M f / c^3)^{1/3}$ and $v_o \to v_{\rm lso} \equiv (\pi G M f_{\rm lso} / c^3)^{1/3}$, where $f$ is the corresponding observed GW frequency and $f_{\rm lso}$ is the same at the last stable orbit given by $c^3/(\pi G M 6^{3/2})$, to obtain
\begin{multline}
    \label{eq: neg_coal_t_circ}
    (t - t_{\rm c})_{\rm LC} = - \frac{5}{256 \eta  v^8}\frac{G M}{c^3} \Biggl[ 1 + z_{\rm L,0} \Biggl\{ \frac{v^8}{\xi } \Biggl(\sin \left(\frac{\xi }{v^8}-\theta_{\rm c}\right) \\ - \sin \left(\frac{\xi }{v_{\rm lso}^8}-\theta_{\rm c}\right) \Biggr) -\frac{8}{3} \cos \left(\frac{\xi }{v^8}-\theta_{\rm c}\right)  \Biggr\} \Biggr]
\end{multline}
where all other terms containing $v_{\rm lso}$, which are just constants, have been absorbed in $t_{\rm c}$. Since the infinitesimal orbital phase $d\phi$ will remain invariant, we can rewrite Equation 3.3a of~\cite{PhysRevD.80.084043} as $d\phi = (v_{\rm o}^3 c^3 / (G M)) (dt_{\rm o} / dv_{\rm o})dv_{\rm o}$. Integrating this in the same limits, we obtain
\begin{equation}
    \label{eq: phi_circ}
    (\phi - \phi_{\rm c})_{\rm LC} = - \frac{1}{32 \eta  v^5} \left[1 - \frac{5}{3}z_{\rm L,0}  \cos \left(\frac{\xi }{v^8}-\theta_{\rm c}\right) \right]
\end{equation}
where $\phi_{\rm c}$ is the orbital phase at the time of coalescence and contains all other constant terms containing $v_{\rm lso}$. 

Finally, we substitute Equations~\eqref{eq: neg_coal_t_circ} and~\eqref{eq: phi_circ} in Equation 14 of~\cite{PhysRevD.80.084043} to obtain the total phase $\Psi_{\rm LC}(f)$ and the phase correction $\Delta \Psi_{\rm LC} (f) = \Psi_{\rm LC}(f) - \Psi(f)$, which is given by
\begin{multline}
    \label{eq: ph_cor_losv_circ}
    \Delta \Psi_{\rm LC} (f) = - \frac{5 z_{\rm L,0}}{128 \eta} \frac{v^3}{\xi} \Biggl[\sin \left(\frac{\xi}{v^8} - \theta_{\rm c} \right) - \sin \left(\frac{\xi}{v_{\rm lso}^8} - \theta_{\rm c} \right)  \Biggr]
\end{multline}
where $\xi \equiv (5/(256 \eta))  (G M \Omega_{\rm det} / c^3)$ and $\Psi(f)$ was provided by the leading-order term of the Equation 18 of~\cite{PhysRevD.80.084043} with $\nu$ replaced by $\eta$. 

To obtain the amplitude correction, we first calculate the new frequency domain amplitude $\mathcal{A}_{\rm LC}$ using the left-hand side of Equation~(4.369) of ~\cite{maggiore2007gravitational} and then calculate $\Delta \mathcal{A}_{\rm LC} / \mathcal{A} \equiv (\mathcal{A}_{\rm LC} - \mathcal{A})/ \mathcal{A}$. Specifically, we account for the Doppler shift in the GW frequency in Equation~(4.29) of~\cite{maggiore2007gravitational} and use $d^2 \Phi / dt_{\rm o}^2 = 2 \pi (df_{\rm o} / dt_{\rm o}) = (6 v_{\rm o}^3 / (G M / c^3)) dv_{\rm o} / dt_{\rm o}$, where $\Phi = 2 \phi$ is the instantaneous GW phase and $dv_{\rm o} / dt_{\rm o}$ is given by Equation~\eqref{eq: dvdt_circ2}, to calculate Equation~(4.361) of the same at the leading order of $(t - t_{\rm c})_{\rm LC}$ given by Equation~\eqref{eq: neg_coal_t_circ}, and obtain:
\begin{equation}
    \label{eq: amp_cor_losv_circ}
    \frac{\Delta \mathcal{A}_{\rm LC}}{\mathcal{A}} = z_{\rm L,0} \Biggl[ \frac{4}{3} \frac{\xi}{v^8} \sin \left(\frac{\xi}{v^8} - \theta_{\rm c} \right) - \frac{1}{6} \cos \left(\frac{\xi}{v^8} - \theta_{\rm c} \right) \Biggr].
\end{equation}

\subsection{Eccentric Outer Orbits}\label{subsec: ecc_der}
For an eccentric outer orbit, the Doppler shift due to the LOSV of the CBC's CoM can be written as
\begin{equation}
    \label{eq: red_ecc}
    z_{\rm LE} = \frac{z_{\rm L,0}  \left[\cos \left(\vartheta+\vartheta_{\rm p}\right) + e_{\rm out} \cos \vartheta_{\rm p}\right]}{\sqrt{1-e_{\rm out}^2}} 
\end{equation}
The Equations~\eqref{eq: dop_shift_circ} retain the same form with $z_{\rm LE}$ in place of $z_{\rm LC}$, while Equation~\eqref{eq: dvdt_circ1} becomes
\begin{multline}
    \label{eq: dvdt_ecc1}
    \frac{dv_{\rm o}}{dt_{\rm o}} =  - \frac{z_{\rm L,0}v_{\rm o}}{3 (1 + z_{\rm LE})^2 \sqrt{1 - e_{\rm out}^2}}\sin(\vartheta + \vartheta_{\rm p}) \frac{d \vartheta}{dt_{\rm u}}  \\ + \left(1 + z_{\rm LE} \right)^{-4/3} \frac{dv_{\rm u}}{dt_{\rm u}}
\end{multline}
Using Equations C8 and C12 of~\cite{Tiwari:2024pvb}, we can write $d\vartheta/dt_{\rm u} = \Omega_{\rm det} (1 + e_{\rm out} \cos \vartheta)^2 / (1 - e_{\rm out}^2)^{3/2}$. Therefore, Equation~\eqref{eq: dvdt_ecc1} takes the form (restricting ourselves to linear order in $z_{\rm L,0}$)
\begin{multline}
    \label{eq: dvdt_ecc2}
    \frac{dv_{\rm o}}{dt_{\rm o}} =  - \frac{z_{\rm L,0} \Omega_{\rm det} v_{\rm o}}{3 (1 - e_{\rm out}^2)^2} (1 + e_{\rm out} \cos \vartheta)^2 \sin(\vartheta + \vartheta_{\rm p})  \\ + \left(1 + z_{\rm LE} \right)^{-4/3} \frac{dv_{\rm u}}{dt_{\rm u}}
\end{multline}
Following the same procedure as in section~\ref{subsec: circ_der}, Equation~\eqref{eq: dvdt_circ2} takes the form
\begin{multline}
    \label{eq: dvdt_ecc3}
    \frac{dv_{\rm o}}{dt_{\rm o}} =  - \frac{z_{\rm L,0} \Omega_{\rm det} v_{\rm o}}{3 (1 - e_{\rm out}^2)^2} (1 + e_{\rm out} \cos \vartheta)^2 \sin(\vartheta + \vartheta_{\rm p})  \\ + \left(1 + \frac{5}{3}\frac{z_{\rm L,0}  \left[\cos \left(\vartheta+\vartheta_{\rm p}\right) + e_{\rm out} \cos \vartheta_{\rm p}\right]}{\sqrt{1-e_{\rm out}^2}} \right) \frac{32 \eta}{5} \frac{c^3 v_{\rm o}^9}{G M} 
\end{multline}
Since, unlike the circular outer orbit scenario, $\vartheta$ does not change linearly with time, we expand $\cos \vartheta$ and $\sin \vartheta$ in $e_{\rm out}$ and the mean anomaly $\zeta \equiv \Omega_{\rm det} (t_{\rm u} - t_{\rm c}) + \theta_{\rm c}$, where $\theta_{\rm c}$ becomes the mean anomaly at the time of coalescence. Specifically, we use Equations 2.84 and 2.85 of~\cite{1999ssd..book.....M}, which are of $\mathcal{O}(e_{\rm out}^5)$ and convergent for $e_{\rm out} \leq 0.6627434$.

Substituting these expansions provided by Equations~\eqref{eq: cos_varth} and~\eqref{eq: sin_varth} of the Appendix~\ref{app: ecc_har} in Equation~\eqref{eq: dvdt_ecc3} and following the same procedure as in section~\ref{subsec: circ_der}, we obtain the phase and amplitude corrections:
\begin{equation}
    \label{eq: ph_cor_losv_ecc}
    \Delta \Psi_{\rm LE} (f) =  - \frac{5 z_{\rm L,0}}{128 \eta \sqrt{1 - e_{\rm out}^2}} \frac{v^3}{\xi} \sum_{n = 0}^4 P_{n} e_{\rm out}^n
\end{equation}
and
\begin{equation}
    \label{eq: amp_cor_losv_ecc}
    \frac{\Delta \mathcal{A}_{\rm LE}}{\mathcal{A}} = \frac{z_{\rm L,0}}{\sqrt{1 - e_{\rm out}^2}} \sum_{n = 0}^4 A_{n} e_{\rm out}^n \,,
\end{equation}
respectively, where $P_n$ are given by Equations~\eqref{eq: P0_ecc}$-$~\eqref{eq: P4_ecc}, while $A_n$ are given by equations~\eqref{eq: A0_ecc}$-$~\eqref{eq: A4_ecc} of the Appendix~\ref{app: ph_amp_corr_eoo}.

\section{Assessing the Measurability of the Parameters}\label{sec: constraints}
To assess the measurability of the mass of the third body and other outer orbit parameters, we first perform the Fisher matrix analysis~\cite{CutlerFlanagan}, invert it to obtain the covariance matrix containing the errors on the LOSV parameters, and finally use the Jacobians to transform it to the covariance matrix in $M_3$ and the size of the outer orbit. We provide the prescription to ensure the validity of the SPA. In addition, we discuss the checks to ensure the stability of the system and discuss other effects such as the gravitational redshift and Shapiro delay as well that can affect the results of this work. In the upcoming sections, we discuss these steps in detail. 

\subsection{Fisher Matrix}\label{subsec: fma}
The Fisher matrix $\bf \Gamma$ for a GW signal incorporating the LOSV effects $h_{\rm TV}(\boldsymbol{\Theta};t)$ can be written as~\cite{CutlerFlanagan}
\begin{equation}
	\label{eq: fm}
    \Gamma_{jk} = \left (\frac{\partial h_{\rm TV}}{\partial \Theta_j} \Bigg \vert \frac{\partial h_{\rm TV}}{\partial \Theta_k} \right) =  4 \mathfrak{R} \int_{f_\mathrm{min}}^{f_\mathrm{max}} \frac{\partial \tilde{h}^\ast_{\rm TV}}{\partial \Theta_j} \frac{\partial \tilde{h}_{\rm TV}}{\partial \Theta_k}   \dfrac{df}{S_n(f)}
\end{equation}
where $\boldsymbol{\Theta}$ is the set of parameters that determine the shape of the signal, while $\Theta_{j,k}$ are individual parameters, and $(|)$ represents a noise-weighted inner product of two time series, $\tilde{}$ denotes the Fourier Transforms, $\ast$ represents the complex conjugate, $f_{\rm min}$ and $f_{\rm max}$ are the minimum and maximum frequencies, respectively, and $S_n(f)$ is the Power Spectral Density (PSD) of the GW detector. 

While calculating the covariance matrix, we rewrite Equation~\eqref{eq: fm} as
\begin{equation*}
    \Gamma_{jk} = 4 \mathfrak{R} \int_{f_\mathrm{min}}^{f_\mathrm{max}}  \left(\frac{\partial \ln \Tilde{h}_{\rm TV}(f)}{\partial \Theta_j}\right)^\ast \frac{\partial \ln \Tilde{h}_{\rm TV}(f)}{\partial \Theta_k}   |\Tilde{h}_{\rm TV}(f)|^2 \frac{df}{S_n(f)}
\end{equation*}
Defining $x \equiv f/f_0$ and $S(x f_0) \equiv S_n(x f_0)/S_0$ to avoid numerical over/under-flows, we can write the above Equation as $\Gamma_{jk} = 4 |\mathcal{A}|^2 \frac{f_0^{-4/3}}{S_0} \Xi_{jk}$ where:
\begin{multline}
    \label{eq: FM_red}
    \Xi_{jk} = \mathfrak{R} \int_{x_\mathrm{min}}^{x_\mathrm{max}}  \left(\frac{\partial \ln \Tilde{h}_{\rm TV}(x f_0)}{\partial \Theta_j}\right)^\ast \frac{\partial \ln \Tilde{h}_{\rm TV}(x f_0)}{\partial \Theta_k} \\ \times  \left(1 + \frac{\Delta \mathcal{A}}{\mathcal{A}} (x f_0)  \right)^2 \frac{x^{-7/3}}{S(x f_0)} dx 
\end{multline}
$x_\mathrm{min} = f_\mathrm{min}/f_0$, and $x_\mathrm{max} = f_\mathrm{max}/f_0$. We can then write the covariance matrix as
\begin{equation}
    \label{eq: CM}
    \boldsymbol{\Sigma} = \boldsymbol{\Gamma}^{-1} = \frac{S_0 f_0^{4/3}}{4 |\mathsf{A}|^2} \boldsymbol{\Xi}^{-1},
\end{equation}
where $\mathcal{A} = \mathsf{A} f^{-7/6}$. We choose $f_0$ to be roughly the frequency at which the detector is most sensitive, and $S_0$  to be the corresponding PSD. However, one can choose these to be any pair of numbers that make the inversion of the matrix $\boldsymbol{\Xi}$ efficient. Similar to~\cite{Tiwari:2024pvb}, we ensure the validity of the inversion of matrix $\boldsymbol{\Xi}$ by the condition ${\rm max}(\vert \boldsymbol{\Xi}^{-1}\boldsymbol{\Xi} - \boldsymbol{I} \vert_{jk} ) \leq 10^{-3}$, where $\boldsymbol{I}$ is the identity matrix.  In addition, we find that the amplitude corrections have a negligible effect on the Fisher matrix. Therefore, we do not include the amplitude correction contributions in the Fisher matrix.

For circular outer orbits, we perform the analysis on the parameter set $\boldsymbol{\Theta}_{\rm C} \equiv \{\ln d_{\rm L},\, \ln \mathcal{M},\, \ln \eta,\, z_{\rm L,0},\, \ln \Omega_{\rm det},\, \theta_{\rm c}\}$, while for eccentric outer orbits we perform the same on $\boldsymbol{\Theta}_{\rm E} \equiv \{\ln d_{\rm L},\, \ln \mathcal{M},\, \ln \eta,\, z_{\rm L,0},\, \ln \Omega_{\rm det},\, e_{\rm out},\, \theta_{\rm c},\, \vartheta_{\rm p}\}$, where $\mathcal{M} \equiv M \eta^{3/5}$ is the cosmologically redshifted chirp mass of the CBC.

\subsection{Validity of the Stationary Phase Approximation and choosing the frequency range}\label{subsec: val_spa}
Because the phase and amplitude corrections presented in this analysis have been derived under the SPA, we ensure that the SPA is valid for the frequency ranges chosen while performing the Fisher matrix analysis. Specifically, we ensure that $df_{\rm o} / dt_{\rm o} > 0$ and choose the minimum frequencies accordingly. We delineate the prescription below.

For COO, using equation~\eqref{eq: dop_shift_circ}, we can write
\begin{equation}
    \label{eq: dfo_by_dto_circ}
    \frac{df_{\rm o}}{dt_{\rm o}} = \frac{1}{\left( 1 + z_{\rm LC} \right)^2} \left[ \frac{df_{\rm u}}{dt_{\rm u}} - \frac{f_{\rm u}}{1 + z_{\rm LC}} \frac{d z_{\rm LC}}{dt_{\rm u}} \right]
\end{equation}
Since the $1 + z_{\rm LC}$ is always $> 0$ for $z_{\rm L,0} \ll 1$, for $df_{\rm o} / dt_{\rm o}$ to be greater than 0, to the leading order in $z_{\rm L,0}$, we require
\begin{equation}
    \label{eq: dfu_by_dtu_crit}
     \frac{df_{\rm u}}{dt_{\rm u}} + f_{\rm u} z_{\rm L,0} \Omega_{\rm det} \sin \left( \Omega_{\rm det} (t_{\rm u} - t_{\rm c}) + \theta_{\rm c} \right) > 0.
\end{equation}
For a slowly varying LOSV such as due to LOSA, this is always satisfied because the first term is always > 0 and the second term is always smaller than the first term due to $z_{\rm L,0} \ll 1$ and $T_{\rm out} \gg t_{\rm obs}$, where $t_{\rm obs}$ is the observation duration and $T_{\rm out}$ is the outer orbital period. In contrast, in the present case, there is no restriction upon $\Omega_{\rm det}$ and hence the second term can even be larger than the first term, and the periodic nature of $z_{\rm LC}$ can lead to $df_{\rm o} / dt_{\rm o} < 0$. Since $\sin \left( \Omega_{\rm det} (t_{\rm u} - t_{\rm c}) + \theta_{\rm c} \right)$ can take values only between $\pm 1$, the most extreme case that could lead to $df_{\rm o} / dt_{\rm o} < 0$ would be $\sin \left( \Omega_{\rm det} (t_{\rm u} - t_{\rm c}) + \theta_{\rm c} \right) = -1$. Therefore, we can write the criterion given by equation~\eqref{eq: dfu_by_dtu_crit} as 
\begin{equation}
    \label{eq: dfu_by_dtu_crit_final}
     \frac{df_{\rm u}}{dt_{\rm u}} > f_{\rm u} z_{\rm L,0} \Omega_{\rm det}.
\end{equation}
Using the relation $df_{\rm u} / dt_{\rm u} = (96 / 5) \left( G \mathcal{M} / c^3 \right)^{5/3} \pi ^{8/3} f_{\rm u}^{11/3}$ in the above equation, we define a critical frequency $f_{\rm SPA,C}$ for the COOs above which the SPA will always be satisfied. This is given by
\begin{equation}
    \label{eq: crit_f_spa_circ}
    f_{\rm SPA,C} = \frac{1}{\pi} \left( \frac{G \mathcal{M}}{c^3} \right)^{-5/8} \left(\frac{5 z_{\rm L,0} \Omega_{\rm det}}{96}  \right)^{3/8}.
\end{equation}

For EOO, equation~\eqref{eq: dfu_by_dtu_crit} takes the form
\begin{equation}
    \label{eq: dfu_by_dtu_crit_ecc}
     \frac{df_{\rm u}}{dt_{\rm u}} + \frac{f_{\rm u} z_{\rm L,0} \Omega_{\rm det} \sin(\vartheta + \vartheta_{\rm p})}{\left( 1 - e_{\rm out}^2 \right)^2} \left( 1 + e_{\rm out} \cos \vartheta  \right)^2  > 0,
\end{equation}
while the criterion given by the equation~\eqref{eq: dfu_by_dtu_crit_final} becomes
\begin{equation}
    \label{eq: dfu_by_dtu_crit_final_ecc}
     \frac{df_{\rm u}}{dt_{\rm u}} > \frac{f_{\rm u} z_{\rm L,0} \Omega_{\rm det}}{\left( 1 - e_{\rm out}^2 \right)^2} \left( 1 + e_{\rm out} \cos \left(\frac{3 \pi}{2} - \vartheta_{\rm p} \right)  \right)^2,
\end{equation}
which has been obtained by fixing $\sin(\vartheta + \vartheta_{\rm p}) = -1$ or equivalently $\vartheta + \vartheta_{\rm p} = 3 \pi / 2$. The critical frequency $f_{\rm SPA,E}$ for the EOOs above which the SPA will always be satisfied is then given by
\begin{equation}
    \label{eq: crit_f_spa_circ}
    f_{\rm SPA,E} = f_{\rm SPA,C} \left( \frac{1 + e_{\rm out} \cos \left(\frac{3 \pi}{2} - \vartheta_{\rm p} \right)}{1 - e_{\rm out}^2} \right)^{3/4}.
\end{equation}

Across all scenarios considered in this {paper}, we assume $4$ years of observation time for DECIGO~\cite{PhysRevD.83.044011, PhysRevD.95.109901} and LISA~\cite{Robson_2019} whenever $f_{\rm SPA,C/E}$ is below their sensitivity band, $[10^{-2}, 10]\, {\rm Hz}$ and $[10^{-4}, 1]\, {\rm Hz}$, respectively. We choose the maximum and minimum frequencies following~\cite{PhysRevD.71.084025}. However, for $f_{\rm SPA,C/E}$ inside their sensitivity band, we choose the minimum frequency to be the maximum of $f_{\rm SPA,C/E}$ and the one obtained following~\cite{PhysRevD.71.084025}. For A+ and ET, we use the frequency band $[{\rm max}(5,\, f_{\rm SPA,C/E}),\,f_{\rm lso}]$ and $[{\rm max}(2,\, f_{\rm SPA,C/E}),\,f_{\rm lso}]$ and use the PSDs provided in Refs.~\cite{A_plus_psd} and~\cite{ET_psd, Hild:2010id}, respectively.

\subsection{Jacobians}\label{subsec: jac_}
Let $\boldsymbol{\Sigma}_{\rm C}$ and $\boldsymbol{\Sigma}_{\rm E}$ be the covariance matrices corresponding to the parameters $\boldsymbol{\Theta}_{\rm C}$ and $\boldsymbol{\Theta}_{\rm E}$, respectively. For getting the errors in the measurement of $M_3$ and $a$, we transform $\boldsymbol{\Sigma}_{\rm C}$ and $\boldsymbol{\Sigma}_{\rm E}$ to the covariance matrix in terms of $M_3$ and $a$. In this section, we calculate the Jacobians of these transformations.

Let $M_{\rm S} = m_{\rm 1,S} + m_{\rm 2,S}$ be the (source frame) total mass of the CBC and $a$ be in units of the Schwarzschild radius of the third body  $R_{\rm s} = 2 G M_3 / c^2$. Then we can write
\begin{equation}
    \label{eq: zl0_fin}
    z_{\rm L,0} = \sqrt{\frac{1}{2a} \frac{M_3}{M_3 + M_{\rm S}}}
\end{equation}
and
\begin{equation}
    \label{eq: om_det_fin}
    \Omega_{\rm det} = \frac{1}{1 + z_{\rm cos}} \frac{c^3}{G M_3} \sqrt{\frac{1}{8a^3} \frac{M_3 + M_{\rm S}}{M_3}}.
\end{equation}
Since these expressions also depend on the total mass of the CBC, we need to account for the contribution due to errors in the measurement of $\mathcal{M}$ and $\eta$ as well while estimating the errors in $M_3$ and $a$. Therefore, we need the Jacobian of the transformation from $(\mathcal{M},\, \eta,\, z_{\rm L,0},\, \Omega_{\rm det}) \to (\mathcal{M},\, \eta,\, M_3,\, a)$, which is given by
\begin{widetext}
\begin{equation}
    \label{eq: jac_inv_fin}
    \boldsymbol{J}^{-1} = \frac{\partial (\mathcal{M}, \eta, M_3, a)}{\partial (\mathcal{M}, \eta, z_{\rm L,0}, \Omega_{\rm det})} = 
    \begin{pmatrix}
        1 & 0 & 0 & 0 \\
        0 & 1 & 0 & 0 \\
        \frac{2 M_3 \eta^{-3/5}}{(M_3 + 3 M_{\rm S}) (1 + z_{\rm cos})} & -\frac{6 M_3 M_{\rm S}}{5 \eta (M_3 + 3 M_{\rm S})} & \frac{3 M_3 (M_3 + M_{\rm S})}{z_{\rm L,0} (M_3 + 3 M_{\rm S})} & -\frac{M_3 (M_3 + M_{\rm S})}{\Omega_{\rm det} (M_3 + 3 M_{\rm S})} \\
        -\frac{a \eta^{-3/5}}{(M_3 + 3 M_{\rm S}) (1 + z_{\rm cos})} & \frac{3 a M_{\rm S}}{5 \eta (M_3 + 3 M_{\rm S})} & - \frac{a (2 M_3 + 3 M_{\rm S})}{z_{\rm L,0} (M_3 + 3 M_{\rm S})} & - \frac{a M_{\rm S}}{\Omega_{\rm det} (M_3 + 3 M_{\rm S})}
    \end{pmatrix}
\end{equation}
\end{widetext}
where $\boldsymbol{J} \equiv \partial (\mathcal{M}, \eta, z_{\rm L,0}, \Omega_{\rm det}) / \partial (\mathcal{M}, \eta, M_3, a)$ is the Jacobian of the transformation from $(\mathcal{M},\, \eta,\, M_3,\, a) \to (\mathcal{M},\, \eta,\, z_{\rm L,0},\, \Omega_{\rm det})$ and is given by Equation~\eqref{eq: jac_fin} of the Appendix~\ref{app: jac_}.

The covariance matrix of $M_3$ and $a$ is then given by $(\boldsymbol{J}^{-1}) (\boldsymbol{J}_{\ln}^{-1}) \boldsymbol{\Sigma}_{\rm C,4} (\boldsymbol{J}_{\ln}^{-1})^{\rm T} (\boldsymbol{J}^{-1})^{\rm T}$ and $(\boldsymbol{J}^{-1}) (\boldsymbol{J}_{\ln}^{-1}) \boldsymbol{\Sigma}_{\rm E,4} (\boldsymbol{J}_{\ln}^{-1})^{\rm T} (\boldsymbol{J}^{-1})^{\rm T}$ for the circular and eccentric outer orbit scenarios, respectively. Here $\boldsymbol{\Sigma}_{\rm C,4}$ and $\boldsymbol{\Sigma}_{\rm E,4}$ are the sub-matrices of $\boldsymbol{\Sigma}_{\rm C}$ and $\boldsymbol{\Sigma}_{\rm E}$ corresponding to the rows $\{\ln \mathcal{M},\, \ln \eta,\, z_{\rm L,0},\, \ln \Omega_{\rm det} \}$, respectively, while $\boldsymbol{J}_{\ln} \equiv \partial (\ln \mathcal{M},\, \ln \eta,\, z_{\rm L,0},\, \ln \Omega_{\rm det}) / \partial (\mathcal{M},\, \eta,\, z_{\rm L,0},\, \Omega_{\rm det})$ is the Jacobian of the transformation from $(\mathcal{M},\, \eta,\, z_{\rm L,0},\, \Omega_{\rm det}) \to (\ln \mathcal{M},\, \ln \eta,\, z_{\rm L,0},\, \ln \Omega_{\rm det})$ and $\boldsymbol{J}_{\ln}^{-1}$ is the same for the transformation $(\ln \mathcal{M},\, \ln \eta,\,, z_{\rm L,0},\, \ln \Omega_{\rm det}) \to (\mathcal{M},\, \eta,\, z_{\rm L,0},\, \Omega_{\rm det})$, which is simply ${\rm diag}(\mathcal{M},\, \eta,\, 1,\, \Omega_{\rm det})$.

For $M_3 \gg M_{\rm S}$, Equation~\eqref{eq: zl0_fin} and~\eqref{eq: om_det_fin} become
\begin{equation}
    \label{eq: zl0_inf}
    z_{\rm L,0} = \frac{1}{\sqrt{2a}},
\end{equation}
\begin{equation}
    \label{eq: om_det_inf}
    \Omega_{\rm det} = \frac{1}{1 + z_{\rm cos}} \frac{1}{\sqrt{8a^3}} \frac{c^3}{G M_3}.
\end{equation}
Therefore, only the submatrix of $\boldsymbol{J}^{-1}$ corresponding to $z_{\rm L,0}$ and $\Omega_{\rm det}$ contributes to the transformation. The Jacobian for the transformation from $(z_{\rm L,0},\, \Omega_{\rm det}) \to (M_3, a)$ is given by
\begin{equation}
    \label{eq: jac_inv_inf}
    \boldsymbol{J}^{-1}_2 =  \frac{\partial (M_3, a)}{\partial (z_{\rm L,0},\, \Omega_{\rm det})} = 
    \begin{pmatrix}
        \frac{3 M_3}{z_{\rm L,0}} & - \frac{M_3}{\Omega_{\rm det}} \\
        - \frac{1}{z_{\rm L,0}^3} & 0  \\
    \end{pmatrix}
\end{equation}
The covariance matrix of $M_3$ and $a$, in this case, is then given by $(\boldsymbol{J}^{-1}_2) (\boldsymbol{J}_{\ln}^{-1}) \boldsymbol{\Sigma}_{\rm C,2} (\boldsymbol{J}_{\ln}^{-1})^{\rm T} (\boldsymbol{J}^{-1}_2)^{\rm T}$ and $(\boldsymbol{J}^{-1}_2) (\boldsymbol{J}_{\ln}^{-1}) \boldsymbol{\Sigma}_{\rm E,2} (\boldsymbol{J}_{\ln}^{-1})^{\rm T} (\boldsymbol{J}^{-1}_2)^{\rm T}$ for the circular and eccentric outer orbit scenarios, respectively. Here $\boldsymbol{\Sigma}_{\rm C,2}$ and $\boldsymbol{\Sigma}_{\rm E,2}$ are the sub-matrices of $\boldsymbol{\Sigma}_{\rm C}$ and $\boldsymbol{\Sigma}_{\rm E}$ corresponding to the rows $\{z_{\rm L,0},\, \ln \Omega_{\rm det} \}$, respectively, while $\boldsymbol{J}_{\ln} \equiv \partial (z_{\rm L,0},\, \ln \Omega_{\rm det}) / \partial (z_{\rm L,0},\, \Omega_{\rm det})$ is the Jacobian of the transformation from $(z_{\rm L,0},\, \Omega_{\rm det}) \to (z_{\rm L,0},\, \ln \Omega_{\rm det})$ and $\boldsymbol{J}_{\ln}^{-1}$ is the same for the transformation $(z_{\rm L,0},\, \ln \Omega_{\rm det}) \to (z_{\rm L,0},\, \Omega_{\rm det})$, which is simply ${\rm diag}(1,\, \Omega_{\rm det})$.

Note that Equations~\eqref{eq: zl0_fin} and~\eqref{eq: zl0_inf} will also have a factor of $\sin \iota_{\rm out}$, which we have fixed to 1 because $\sin \iota_{\rm out}$ is degenerate with $M_3$ and $a$, and this degeneracy cannot be broken. As a result, the errors in $M_3$ and $a$ should be treated as lower limits.

\subsection{Stability of the system}\label{subsec: stab}
Let $q_{\rm out} = M_3 / M_{\rm S}$ be the mass ratio of the triple, $a_{\rm in}$ be the size of the inner orbit, and $\iota_{\rm mut}$ be the mutual inclination of the inner and outer orbit. Then the stability criteria for the third body not to escape ---  Equation (2) of~\cite{2022MNRAS.516.4146V}, which is Equation (90) of~\cite{2001MNRAS.321..398M} --- can be written as 
\begin{equation}
    \label{eq: stab}
    \frac{a_{\rm crit}}{a_{\rm in}} = \frac{2.8}{R_{\rm s}} \left[ \frac{(1 + q_{\rm out}) (1 + e_{\rm out})}{\sqrt{1 - e_{\rm out}}} \right]^{2/5} \left(1 - 0.3 \frac{\iota_{\rm mut}}{\pi} \right)
\end{equation}
where $0 \leq \iota_{\rm mut} \leq \pi$ and $R_{\rm s}$ in the denominator on the right-hand side is present because we are taking $a_{\rm crit}$ to be in units of $R_{\rm s}$. Throughout this {paper}, we fix $\iota_{\rm mut} = \pi / 2$ because we assume the inner orbit is face-on and the outer orbit is edge-on. The systems with $a \geq a_{\rm crit}$ for circular outer orbits and $a (1 - e_{\rm out}) \geq a_{\rm crit}$ for eccentric outer orbits are deemed {\it stable}. Using equation~\eqref{eq: stab} and $a_{\rm in} = (G M_{\rm S} / (\pi^2 f_{\rm S}^2))^{1/3}$, for $\iota_{\rm mut} = \pi / 2$, we can write
\begin{multline}
    \label{eq: stab_rs}
    a_{\rm crit} = 1.94 \times 10^3 \left[ \frac{1 + e_{\rm out}}{\sqrt{1 - e_{\rm out}}} \right]^{2/5} \left( \frac{1 + q_{\rm out}}{1.345} \right)^{2/5} \\ \times \left( \frac{M_{\rm S}}{2.9} \right)^{1/3} \frac{1}{M_3} \left( \frac{2}{f_{\rm S}} \right)^{2/3},
\end{multline}
where $f_{\rm S} = f_{\rm u} (1 + z_{\rm cos})$ is the source frame GW frequency.

For the CBC not to be tidally disrupted by the tertiary, the critical separation between the CBC and the tertiary is given by\footnote{This can be obtained by equating the tidal force on the CBC due to the tertiary to the CBC's internal force. Specifically, it can be obtained by $G M_{\rm S}/a_{\rm in}^2 = G M_3 a_{\rm in} / a_{\rm crit,dis}^3$.} $a_{\rm crit,dis} = a_{\rm in} q_{\rm out}^{1/3} / R_{\rm s}$. If $a > a_{\rm crit,dis}$ for circular outer orbits and $a (1 - e_{\rm out}) > a_{\rm crit,dis}$ for the eccentric outer orbits, the CBC will not be tidally disrupted. Using Equation~\eqref{eq: stab} and the above expression for $a_{\rm crit,dis}$ we can write  
\begin{equation}
    \label{eq: esc_vs_tid}
    \frac{a_{\rm crit}}{a_{\rm crit,dis}} = 2.8 \left(\frac{1 + q_{\rm out}}{q_{\rm out}^{5/6}} \frac{1 + e_{\rm out}}{\sqrt{1 - e_{\rm out}}} \right)^{2/5} \left(1 - 0.3 \frac{\iota_{\rm mut}}{\pi} \right).
\end{equation}
It can be seen that for $q_{\rm out} \ll 1$, $a_{\rm crit} / a_{\rm crit,dis} \propto q_{\rm out}^{-1/3} > 1$, and $q_{\rm out} \gg 1$, $a_{\rm crit} / a_{\rm crit,dis} \propto q_{\rm out}^{1/15} > 1$. Therefore, any system considered in this analysis that is stable against escape will not be tidally disrupted. As a result, we will demarcate only the regions where the system is unstable.

\subsection{Gravitational Redshift}\label{app: grav_red}
Similar to the Doppler shift, the gravitational redshift can also lead to a change in GW frequency and to a time dilation. For COOs, it will just be a constant because $a$ will be constant and hence will be degenerate with the masses of the CBC. However, in the case of EOOs, it can lead to modulations. Specifically, in the EOO's case, the gravitational redshift is given by~\cite{Meiron:2016ipr}
\begin{equation}
    \label{eq: grav_red_def}
    z_{\rm G} = \frac{1}{a} \frac{1 + e_{\rm out} \cos \vartheta}{1 - e_{\rm out}^2},
\end{equation}
where $a$, as usual, is in units of $R_{\rm s}$. Using equation~\eqref{eq: zl0_fin}, we can rewrite the equation~\eqref{eq: grav_red_def} as
\begin{equation}
    \label{eq: grav_red}
    z_{\rm G} = 2 \left(\frac{z_{\rm L,0}}{\sqrt{1 - e_{\rm out}^2}} \right)^2 \left( 1 + \frac{M_{\rm S}}{M_3} \right) (1 + e_{\rm out} \cos \vartheta).
\end{equation}
From equations~\eqref{eq: red_ecc} and~\eqref{eq: grav_red}, we see that $z_{\rm G} / z_{\rm LE} \propto \left(z_{\rm L,0} / \sqrt{1 - e_{\rm out}^2} \right) (1 + M_{\rm S} / M_3)$. For $M_3 \gg M_{\rm S}$, $1 + M_{\rm S} / M_3 \approx 1$, while for the smallest value of $M_3 = 1 \, M_{\odot}$ paired with a CBC of $M_{\rm S} = 20 \, M_{\odot}$ --- most massive system with which it has been paired (see Table~\ref{tab: tab_sys}), $1 + M_{\rm S} / M_3 = 21$. Since we are in $z_{\rm L,0} / \sqrt{1 - e_{\rm out}^2} \ll 1$ regime, for the most extreme case $z_{\rm L,0} / \sqrt{1 - e_{\rm out}^2} = 0.05$, we find that $z_{\rm G} / z_{\rm LE} \propto (1 + M_{\rm S} / M_3) \left(z_{\rm L,0} / \sqrt{1 - e_{\rm out}^2} \right) = 1.05$, which is already ruled out (see Figure~\ref{fig: Ap_ET_BNS_BBH_SBH_IMBH_EO}) due to stability criteria discussed in the Section~\ref{subsec: stab}. Therefore, in the parameter space of interest, the gravitational redshift will be a subdominant effect in comparison to the Doppler shift.

\subsection{Shapiro Delay}\label{app: shap}
Apart from the Doppler shift and the gravitational redshift, another interesting effect is the Shapiro delay, which causes a shift in the arrival time of GWs. This time delay can be written as~\cite{Meiron:2016ipr, 1986ARA&A..24..537B}
\begin{equation}
    \label{eq: time_shift_shap}
    \Delta t_{\rm SE} = \frac{2 G M_3}{c^3} \ln \left \vert \frac{1 + e_{\rm out} \cos \vartheta}{1 - \sin \iota_{\rm out} \cos (\vartheta + \vartheta_{\rm p})}  \right \vert,
\end{equation}
where $\rm E$ stands for EOO. Because $d\Phi/dt = 2 \pi f$ and $d\Phi$ remains invariant, a time delay $t_{\rm u} \to t_{\rm u} + \Delta t_{\rm SE}$ will also shift the GW frequency, which is given by $f_{\rm o} = f_{\rm u}\,dt_{\rm u} / d( t_{\rm u} + \Delta t_{\rm SE}) = f_{\rm u} / (1 + d \Delta t_{\rm SE}/dt_{\rm u})$. Therefore, the term $d \Delta t_{\rm SE}/dt_{\rm u}$ has an effect similar to a Doppler shift and is given by
\begin{multline}
    \label{eq: dts_by_dtu}
    \frac{d \Delta t_{\rm SE}}{dt_{\rm u}} = \frac{2 G M_3}{c^3} \frac{\Omega_{\rm det}  (1 + e_{\rm out} \cos \vartheta)}{\left(1-e_{\rm out}^2 \right)^{3/2}} \\ \times \frac{\left[\sin \iota_{\rm out} \left(e_{\rm out} \sin \vartheta_{\rm p} + \sin (\vartheta+\vartheta_{\rm p}) \right)+e_{\rm out} \sin \vartheta \right]}{\left(\sin \iota_{\rm out} \cos (\vartheta+\vartheta_{\rm p}) - 1 \right)},
\end{multline}
where we have used the expression of $d\vartheta/dt_{\rm u}$ used in the section~\ref{subsec: ecc_der}. Using equations~\eqref{eq: zl0_fin} and~\eqref{eq: om_det_fin}, we can write $\Omega_{\rm det} G M_3 / c^2 = (1 + M_{\rm S} / M_3)^2 \, z_{\rm L,0}^3 / (1 + z_{\rm cos}) $. Therefore, we can write
\begin{equation}
    \label{eq: dts_by_dt_zlo}
    \frac{d \Delta t_{\rm SE}}{dt_{\rm u}} = \left(\frac{z_{\rm L,0}}{\sqrt{1 - e_{\rm out}^2}} \right)^3 \left(1 + \frac{M_{\rm S}}{M_3} \right)^2 \frac{2}{1 + z_{\rm cos}} F_{\rm E} (\vartheta)
\end{equation}
where
\begin{multline}
    \label{eq: F_th_def}
    F_{\rm E} (\vartheta) = \frac{(1 + e_{\rm out} \cos \vartheta)}{(\sin \iota_{\rm out} \cos (\vartheta+\vartheta_{\rm p})-1)} \\ \times \left[\sin \iota_{\rm out} \left (e_{\rm out} \sin \vartheta_{\rm p} + \sin (\vartheta+\vartheta_{\rm p}) \right)+e_{\rm out} \sin \vartheta \right].
\end{multline}
Notice that $d \Delta t_{\rm SE}/dt_{\rm u} \propto \left(z_{\rm L,0} / \sqrt{1 - e_{\rm out}^2} \right)^3$, while $z_{\rm LE} \propto z_{\rm L,0} / \sqrt{1 - e_{\rm out}^2}$. For $M_3 \gg M_{\rm S}$, $(1 + M_{\rm S} / M_3)^2 \approx 1$, while for smallest value of $M_3 = 1 \, M_{\odot}$ considered in this work corresponding to a $M_{\rm S} = 20 \, M_{\odot}$, $(1 + M_{\rm S} / M_3)^2 = 441$. Since we are in $z_{\rm L,0} / \sqrt{1 - e_{\rm out}^2} \ll 1$ regime, for the most extreme case of $z_{\rm L,0} / \sqrt{1 - e_{\rm out}^2} = 0.05$, we find that $(d \Delta t_{\rm SE}/dt_{\rm u}) / z_{\rm LE} \propto (1 + M_{\rm S} / M_3)^2 \left(z_{\rm L,0} / \sqrt{1 - e_{\rm out}^2} \right)^2 = 1.1025$, which is already ruled out (see Figures~\ref{fig: Ap_ET_BNS_BBH_SBH_IMBH_CO} and~\ref{fig: Ap_ET_BNS_BBH_SBH_IMBH_EO}) due to stability criteria discussed in the Section~\ref{subsec: stab}. 
Note that for some values of $\vartheta$, the $F_{\rm E}(\vartheta)$ can still become very large. To get a qualitative idea of how large this term can be, we consider the COO and set $e_{\rm out} = 0$ and $\vartheta_{\rm p} = \theta_{\rm c}$ in equation~\eqref{eq: F_th_def} and take its derivative with respect to $\vartheta$ and obtain
\begin{equation}
    \label{eq: F_der_th}
    \frac{d F_{\rm C}}{d \vartheta} = \frac{\sin \iota_{\rm out} [\sin \iota_{\rm out} - \cos (\vartheta + \theta_{\rm c})]}{(1 - \sin \iota_{\rm out} \cos (\vartheta + \theta_{\rm c}))^2}.
\end{equation}
Setting $dF_{\rm C}/d\vartheta = 0$ gives us $\cos (\vartheta + \theta_{\rm c}) = \sin \iota_{\rm out}$ and $\sin (\vartheta + \theta_{\rm c}) = \pm \cos \iota_{\rm out}$. The second derivative of $F_{\rm C}$ is given by
\begin{multline}
    \label{eq: F_sec_der_th}
    \frac{d^2 F_{\rm C}}{d \vartheta^2} = \sin (\vartheta + \theta_{\rm c}) \sin \iota_{\rm out}  \\ \times \frac{\left( \sin\iota_{\rm out} \cos (\vartheta + \theta_{\rm c})-2 \sin ^2 \iota_{\rm out}+1 \right)}{(1 - \sin \iota_{\rm out} \cos (\vartheta + \theta_{\rm c}))^3}.
\end{multline}
For $\cos (\vartheta + \theta_{\rm c}) = \sin \iota_{\rm out}$ and $\sin (\vartheta + \theta_{\rm c}) = - \cos \iota_{\rm out}$, we can see that $d^2 F_{\rm C} / d \vartheta^2 = - \sin \iota_{\rm out} / \cos^3 \iota_{\rm out} < 0$ whenever $\iota_{\rm out} \leq \pi / 2$, which indicates that this point is a maximum with $F_{\rm C,max} = \tan \iota_{\rm out}$. For $\iota_{\rm out} = \pi / 2$, this becomes $\infty$. However, changing $\iota_{\rm out}$ to $89^\circ$ would make
$F_{\rm C,max} \approx 57.3$ and $(d \Delta t_{\rm SC} / dt_{\rm u})_{\rm max} \approx 0.013 (1 + M_{\rm S} / M_3)^2$ for $z_{\rm L,0} = 0.05$ and $z_{\rm cos} = 0.105$, where $d \Delta t_{\rm SC} / dt_{\rm u}$ has been obtained by setting $e_{\rm out} = 0$ and $\vartheta_{\rm p} = \theta_{\rm c}$ in equation~\eqref{eq: dts_by_dt_zlo}. For $M_{\rm S} = 20 \, M_{\odot}$, the maximum tertiary mass that is not ruled out due to the stability criteria discussed in Section~\ref{subsec: stab} is $M_3 \sim 2\, M_{\odot}$ (see Figure~\ref{fig: Ap_ET_BNS_BBH_SBH_IMBH_CO}), which gives $(d \Delta t_{\rm SC} / dt_{\rm u})_{\rm max} = 1.57 > z_{\rm L,0}$. For a more massive tertiary such as $M_3 = 25 \, M_{\odot}$, the same becomes $(d \Delta t_{\rm SC} / dt_{\rm u})_{\rm max} = 0.042 < z_{\rm L,0}$. Therefore, for outer orbits that are close to edge-on and $M_3 \lesssim M_{\rm S}$, the Shapiro delay can become dominant, and a study of both effects combined would be essential. We plan to investigate this in more detail in future work. 

\begin{table*}[ht!]
    \centering
    \begin{tabular}{|c|c|c|c|c|c|}
        \hline
        \textbf{Detector: CBC System} & $m_{\rm 1,S} - m_{\rm 2,S} \, [M_{\odot}]$ & $d_{\rm L}$ [Gpc] & $z_{\rm cos}$ & \textbf{SBH - IMBH} & \textbf{SMBH} \\
        \hline
        \hline
        A+: BNS & $1.6-1.3$ & 0.1 & 0.022 & COO \& EOO & COO \\
        \hline
        A+: NSBH & $5-1.4$ & 0.1 & 0.022 & - & COO \\
        \hline
        A+: BBH & $10-10$ & 0.5 & 0.105 & COO \& EOO &  - \\
        \hline
        ET: BNS & $1.6-1.3$ & 0.1 & 0.022 & COO \& EOO &  COO \& EOO \\
        \hline
        ET: BBH & $10-10$ & 1 & 0.198 & COO \& EOO & COO \\
        \hline
        ET: BBH2 & $30-30$ & 0.1 & 0.022 & - & COO \\
        \hline
        DECIGO: BBH & $100-100$ & 1 & 0.198 & - & COO \& EOO \\
        \hline
        LISA: BBH & $100-100$ & 1 & 0.198 & - & COO \& EOO \\
        \hline
    \end{tabular}
    \caption{Table of system and detector configurations considered in this {paper}. COO and EOO refer to circular and eccentric outer orbits, respectively, while SBH, IMBH, and SMBH refer to stellar mass BH, intermediate-mass BH, and supermassive BH, respectively, as the third body. We have used Planck18 cosmology~\cite{Planck:2018vyg} to convert $d_{\rm L}$ to $z_{\rm cos}$.}
    \label{tab: tab_sys}
\end{table*}

\subsection{Dynamical Effects}\label{sec: dyn_eff}
While the Doppler shift, gravitational redshift, and Shapiro delay are extrinsic to the CBC because they do not alter the physical inspiral of the CBC, the effects such as the nodal precession, Kozai-Lidov effect, and tidal force effects lead to changes in the dynamics of CBC and hence can change the shape and evolution of the inner orbit of the CBC\footnote{Note that since there is no analytical frequency-domain waveform which accounts for the tidal dephasing valid for all values of $T_{\rm out}$, we do not demarcate these regions because it is out of scope of this work.}. From equation (30) of \cite{Meiron:2016ipr} (see also~\cite{2013MNRAS.431.2155N, Naoz:2012bx}), we see that the nodal precession rate to leading quadrupole order is $\propto \iota_{\rm mut}$. Since we have $\iota_{\rm mut} = \pi/2$, nodal precession will be absent. However, this can also lead to excitation of large eccentricity of the inner orbit and to inclination flips of the orbits due to the Kozai-Lidov effect. The maximum eccentricity excited is given by 
\begin{equation}
    \label{eq: ecc_ex}
    e_{\rm in, max} = \sqrt{1 - \frac{5}{3} \cos^2 \iota_{\rm mut}},
\end{equation}
which is equation (20) of~\cite{2016ARA&A..54..441N} and gives $e_{\rm in,max} = 1$ for $\iota_{\rm mut} = \pi/2$. These changes happen over a time-scale~\cite{2016ARA&A..54..441N} 
\begin{equation}
    \label{eq: t_kl}
    t_{\rm KL} \sim \frac{8}{15 \pi} \left( 1 + \frac{M_{\rm S}}{M_3} \right) \frac{f_{\rm S} T_{\rm out}^2}{2} (1 - e_{\rm out}^2)^{3/2},
\end{equation}
where we have used the relation $T_{\rm in} = 2/f_{\rm S}$. However, the precession of the inner orbit due to GR, which happens over a time-scale~\cite{2016ARA&A..54..441N}
\begin{equation}
    \label{eq: t_prec}
    t_{\rm prec,GR} \sim 2 \pi \frac{a_{\rm in}^{5/2} c^2 (1 - e_{\rm in}^2)}{3 G^{3/2} M_{\rm S}^{3/2}},
\end{equation}
where $e_{\rm in}$ is the eccentricity of the inner orbit, can suppress these effects. Specifically, when $t_{\rm prec,GR} < t_{\rm KL}$, eccentricity excitations as well as the inclination flips will be suppressed~\cite{2016ARA&A..54..441N}. From equations~\eqref{eq: t_kl} and~\eqref{eq: t_prec}, we can write 
\begin{equation}
    \label{eq: tkl_by_tprec}
    \frac{t_{\rm prec,GR}}{t_{\rm KL}} = \frac{5 c^8 (1 - e_{\rm in}^2)}{64 \pi ^{5/3} a^3 \left(1-e_{\rm out}^2\right)^{3/2} G^{8/3} M_3^2 M_{\rm S}^{2/3} f_{\rm S}^{8/3}}.
\end{equation}
For quasi-circular CBCs, we can write
\begin{multline}
    \label{eq: tkl_by_tprec_qc}
    \left(\frac{t_{\rm prec,GR}}{t_{\rm KL}} \right)_{\rm QC} = 127.84 \frac{1}{\left(1-e_{\rm out}^2\right)^{3/2}} \left( \frac{2}{f_{\rm S}} \right)^{8/3} \\ \times  \left( \frac{2.9}{M_{\rm S}} \right)^{2/3} \left( \frac{1}{M_3} \right)^2 \left( \frac{10^3}{a} \right)^3,
\end{multline}
where the masses are in units of $M_{\odot}$, GW frequency is in Hz, and $a$ is in $R_{\rm s}$. Using $t_{\rm prec,GR} / t_{\rm KL} = 1$, we define a critical distance $a_{\rm crit,KL}$ above which the eccentricity excitations and inclination flips will be suppressed due to GR precession. This is given by

\begin{figure*}[ht!]
    \centering
    \includegraphics[width=0.975\linewidth]{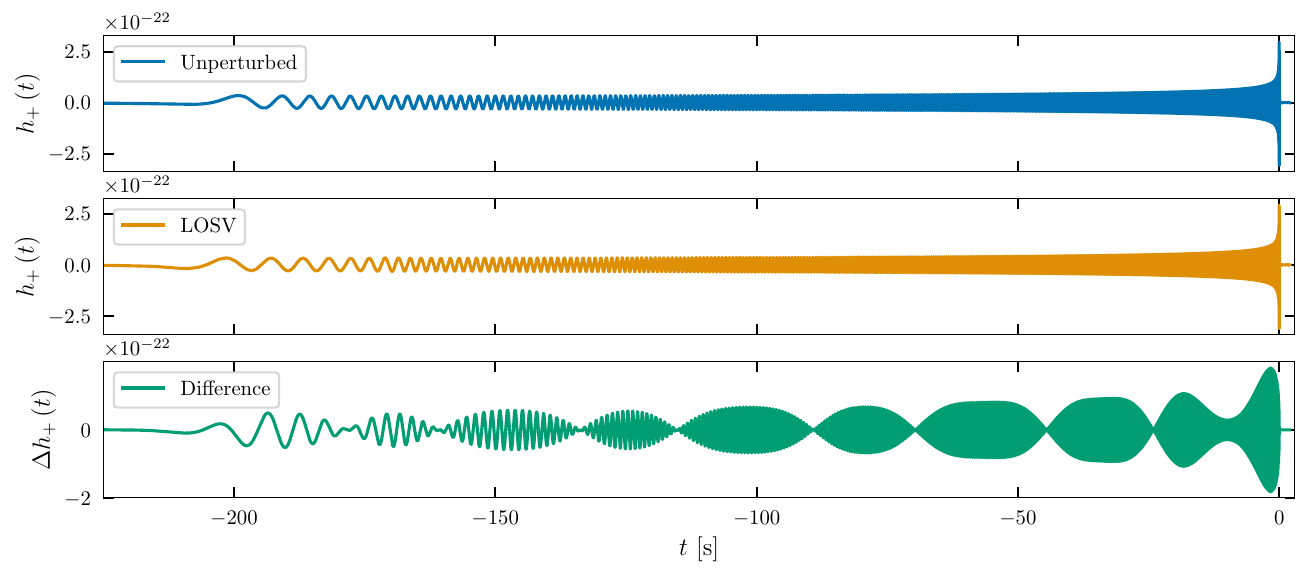}
    \caption{\textbf{Example Waveform}: The {\it top panel} shows the time domain waveform of a non-spinning static BBH at 500 Mpc having component masses $m_{\rm 1,S} = m_{\rm 2,S} = 10 \, M_{\odot}$, the {\it middle panel} shows the same when there is a $8 \, M_{\odot}$ BH in the vicinity of this BBH at $2.25 \times 10^3 \, R_{\rm s}$ in a COO perturbing the motion of its CoM --- this configuration leads to $z_{\rm L,0} = 8 \times 10^{-3}$ and $\Omega_{\rm det} = 0.142 \, \rm Hz$, and the {\it bottom panel} shows the difference between the two waveforms.} 
    \label{fig: waveform_ex}
\end{figure*}

\begin{equation}
    \label{eq: acrit_kl_def}
    a_{\rm crit,KL} = \frac{5.04 \times 10^3 }{\left(1-e_{\rm out}^2\right)^{1/2}} \left( \frac{2}{f_{\rm S}} \right)^{8/9}  \left( \frac{2.9}{M_{\rm S}} \right)^{2/9} \left( \frac{1}{M_3} \right)^{2/3}.
\end{equation}
Using equations~\eqref{eq: stab_rs} and~\eqref{eq: acrit_kl_def}, we can write
\begin{multline}
    \label{eq: acrit_kl_acrit_comp}
    \frac{a_{\rm crit,KL}}{a_{\rm crit}} = 2.598 \left[ \frac{\sqrt{1 - e_{\rm out}}}{1 + e_{\rm out}} \right]^{2/5} \frac{1}{\left(1-e_{\rm out}^2\right)^{1/2}} \\ \times \left( \frac{2}{f_{\rm S}} \right)^{2/9}  \left( \frac{2.9}{M_{\rm S}} \right)^{5/9} M_3^{1/3} \left( \frac{1.345}{1 + q_{\rm out}} \right)^{2/5}.
\end{multline}
Since $a_{\rm crit,KL} / a_{\rm crit}$ depends on the tertiary mass through $q_{\rm out}$ as well, we find no monotonic trend in favor of $a_{\rm crit,KL}$ or $a_{\rm crit}$ for a fixed $f_{\rm S}$ (see Figure~\ref{fig: acrit_comp} of Appendix~\ref{app: add_fig}). However, we do find that for most parts of the parameter space considered in this work, the more stringent criteria on $a$ are given by $a_{\rm crit}$, which is because of $a_{\rm crit,KL} / a_{\rm crit}$ being $\propto f_{\rm S}^{-2/9}$ and large $f_{\rm S}$ being chosen to ensure the validity of the SPA discussed in Section~\ref{subsec: val_spa}. Therefore, while we demarcate the regions where the system is unstable against escape, we do not demarcate $a < a_{\rm crit,KL}$ because it falls below $200 \, R_{\rm s}$, which is the minimum size of the outer orbit considered in this analysis. 

\begin{figure*}[ht!]
    \centering
    \includegraphics[width=0.495\linewidth]{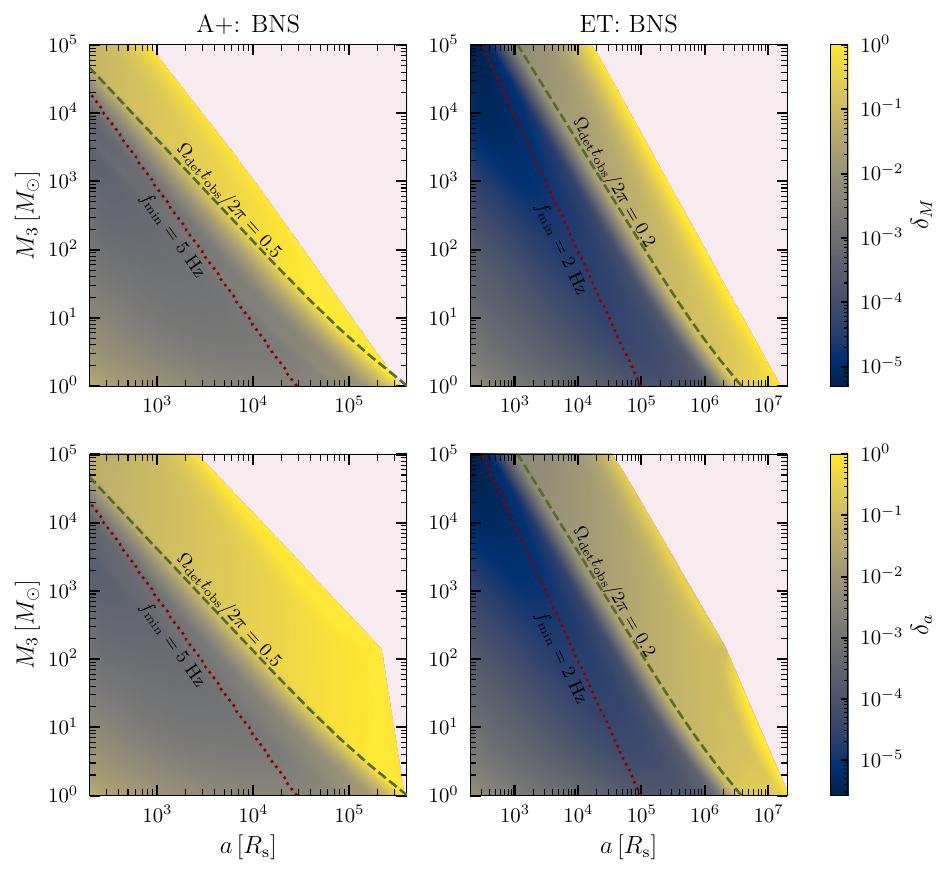}
    \includegraphics[width=0.495\linewidth]{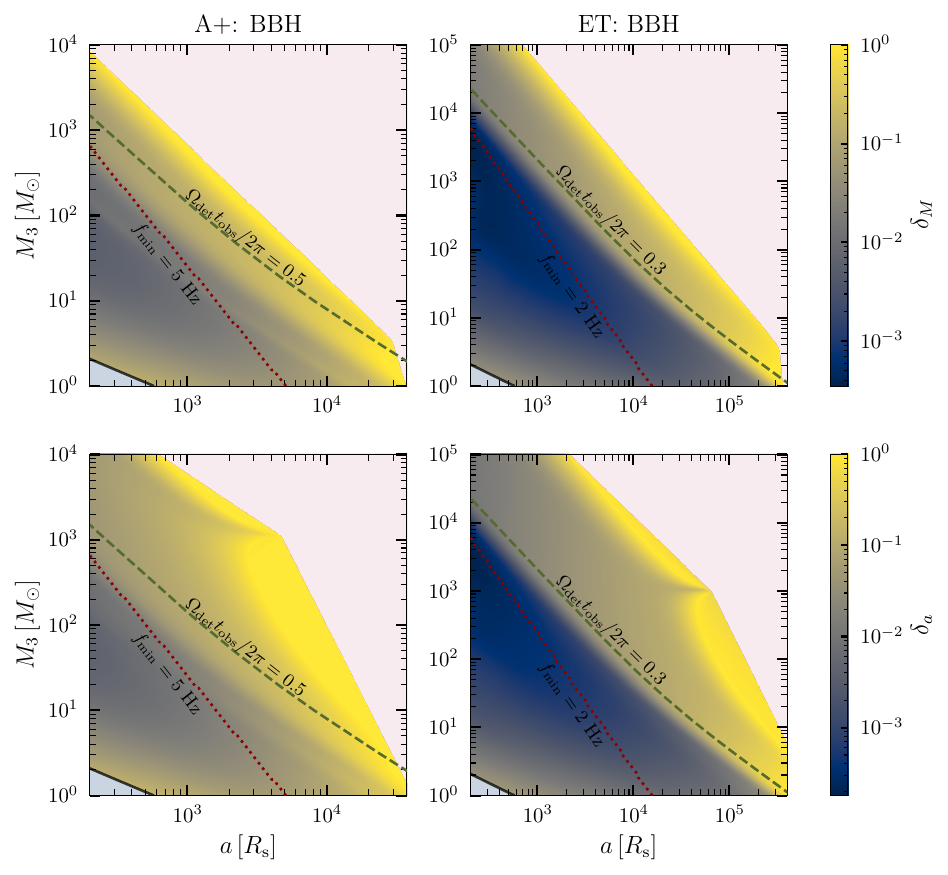}
    \caption{\textbf{SBH-IMBH:} {\it Left two panels} show the relative errors in the measurement of mass of the tertiary $M_3$ ({\it top panels}) and radius of the outer orbit $a$ ({\it bottom panels}) over a grid of $M_3$ and $a$ for the A+: BNS and ET: BNS cases mentioned in Table~\ref{tab: tab_sys} in COO scenario, while the {\it right two panels} show the same for A+: BBH and ET: BBH cases. The patches on the upper right represent the parameter space where either $\delta_X > 1$ for parameter $X$ or the Fisher matrix inversion becomes inefficient, while the patches on the bottom left, demarcated by the solid line, represent the region of parameter space where the three-body system is unstable against escape. The dashed lines represent the contours of a constant $\Omega_{\rm det} t_{\rm obs} / 2 \pi$ demarcating the $\Omega_{\rm det} t_{\rm obs} / 2 \pi \ll 1$ region (upper right to the line) and dotted lines represent the contours of a constant $f_{\rm min}$ and demarcate the region (lower left to the line) in which the minimum frequency is chosen following the prescription delineated in Section~\ref{subsec: val_spa} to ensure the validity of the SPA.}
    \label{fig: Ap_ET_BNS_BBH_SBH_IMBH_CO}
\end{figure*}

\subsection{Signal-to-Noise Ratio (SNR)}\label{sec: SNR_cal}
The optimal SNR for the GW signal~\eqref{eq: WF}, after neglecting the amplitude corrections, is given by
\begin{equation}
\varrho = \sqrt{4 |\mathsf{A}|^2 \mathcal \int_{f_{\rm min}}^{f_{\rm max}} \frac{f^{-7/3}}{S_n(f)} df}
\end{equation}
Throughout this {paper}, we assume the CBCs are face-on, which maximizes the SNR. However, when the CBC is inclined, the SNRs will be reduced by a factor of $\frac{\mathcal{Q} (\iota)}{\mathcal{Q} (\iota = 0)}$ where
\begin{equation}
\mathcal{Q}(\iota) = \sqrt{\left( \frac{1 + \cos^2\iota}{2}\right)^2 + \cos^2\iota}   
\end{equation}
and $\iota$ is the inclination of the CBC relative to the LOS \cite{Robson_2019}. We set a minimum threshold of $\varrho = 4$ for a CBC to be detectable.

\section{Results}\label{sec: results}
Before assessing the measurability of the parameters pertaining to the outer orbit and mass of the tertiary, in Figure~\ref{fig: waveform_ex}, we show the time domain waveforms of a $10-10 \, M_{\odot}$ non-spinning BBH merger at 500 Mpc generated using the inverse Fourier Transform of the frequency domain waveform generated using \textsc{IMRPhenomXP}~\cite{Pratten:2020ceb} implemented in \textsc{LALSuite}~\cite{2020ascl.soft12021L} as the base waveform in the frequency range $5 \, {\rm Hz} - f_{\rm lso}$. The {\it top panel} shows the waveform when this BBH is static, the {\it middle panel} shows the same when its CoM is perturbed by a third BH of mass $8 \, M_{\odot}$ at $2.25 \times 10^3 \, R_{\rm s}$ to move in a circular orbit around the system's barycenter\footnote{We have checked that this system is stable against escape.} with $\theta_{\rm c} = 0.1$ radians, and the {\it bottom panel} shows the difference between the perturbed and unperturbed waveforms. We notice that both waveforms go in and out of phase repeatedly, which is because the outer orbital period (44.1 s) is smaller than the signal duration of the BBH (203.5 s), and as a result the time delay and frequency shift will periodically drop to zero.
We also show the waveforms for the same system when the outer orbit is eccentric in Figure~\ref{fig: waveform_ex_eo} of the Appendix~\ref{app: add_fig_wav} and the comparison between the two perturbed waveforms, COO and EOO, in Figure~\ref{fig: waveform_ex_co_eo_diff} of the same. In addition, we also compute the {\it match}, maximized over $t_{\rm c}$ and $\phi_{\rm c}$, between the unperturbed and perturbed waveforms in A+ using \textsc{PyCBC}~\cite{alex_nitz_2024_10473621} and find it to be 0.76, which is much smaller than the typical acceptable threshold of 0.97 for the match between two waveforms to label them as similar. This is also the minimal match criterion for standard CBC template banks.

\begin{figure}[ht!]
    \centering
    \includegraphics[width=0.975\linewidth]{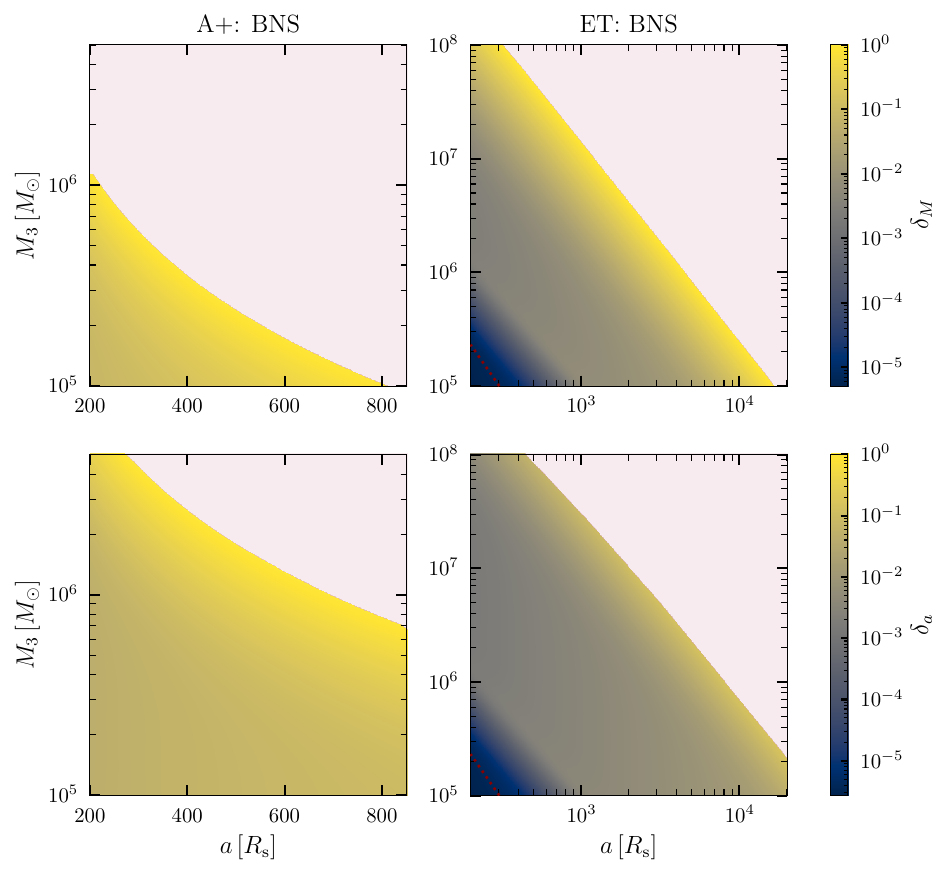}
    \caption{\textbf{SMBH:} The {\it left} and {\it right panels} show the relative errors in the measurement of $M_3$ ({\it top panels}) and $a$ ({\it bottom panels}) over a grid of $M_3$ and $a$ for the A+: BNS and ET: BNS cases mentioned in Table~\ref{tab: tab_sys} in COO scenario, respectively. The patches on the upper right and the dotted lines have the same meaning as in Figure~\ref{fig: Ap_ET_BNS_BBH_SBH_IMBH_CO}.}
    \label{fig: Ap_ET_bns_SMBH}
\end{figure}

\begin{figure*}[!ht]
    \centering
    \includegraphics[width=0.495\linewidth]{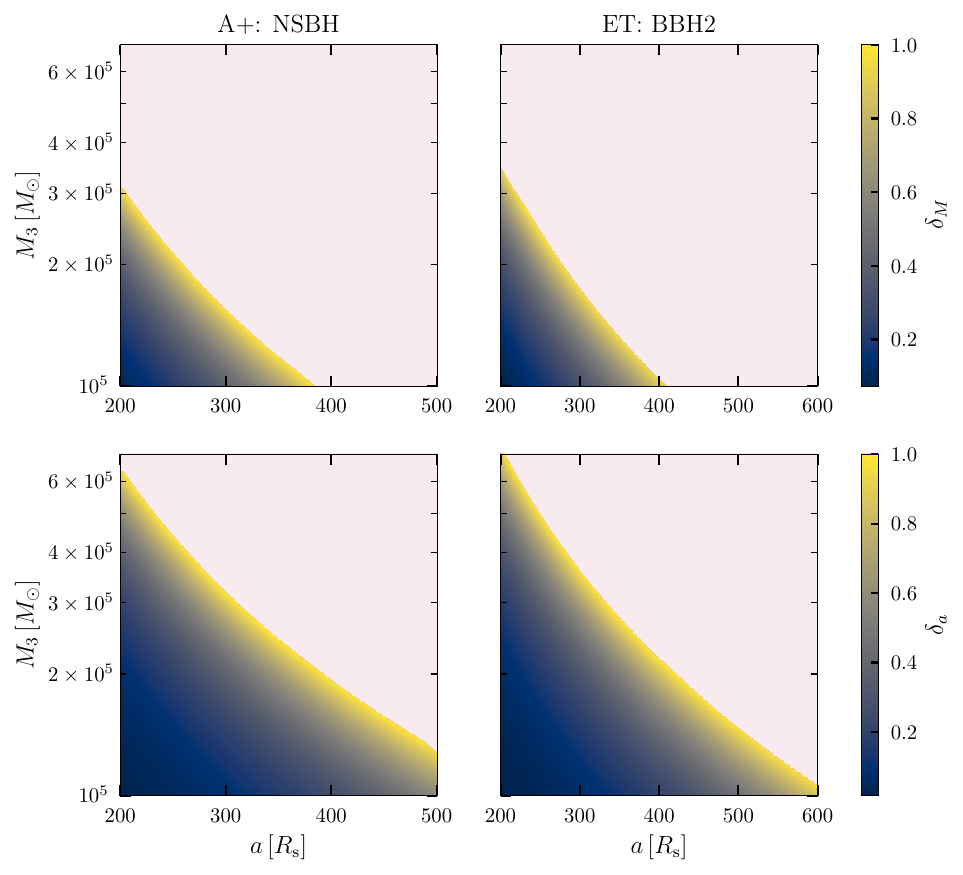}
    \includegraphics[width=0.495\linewidth]{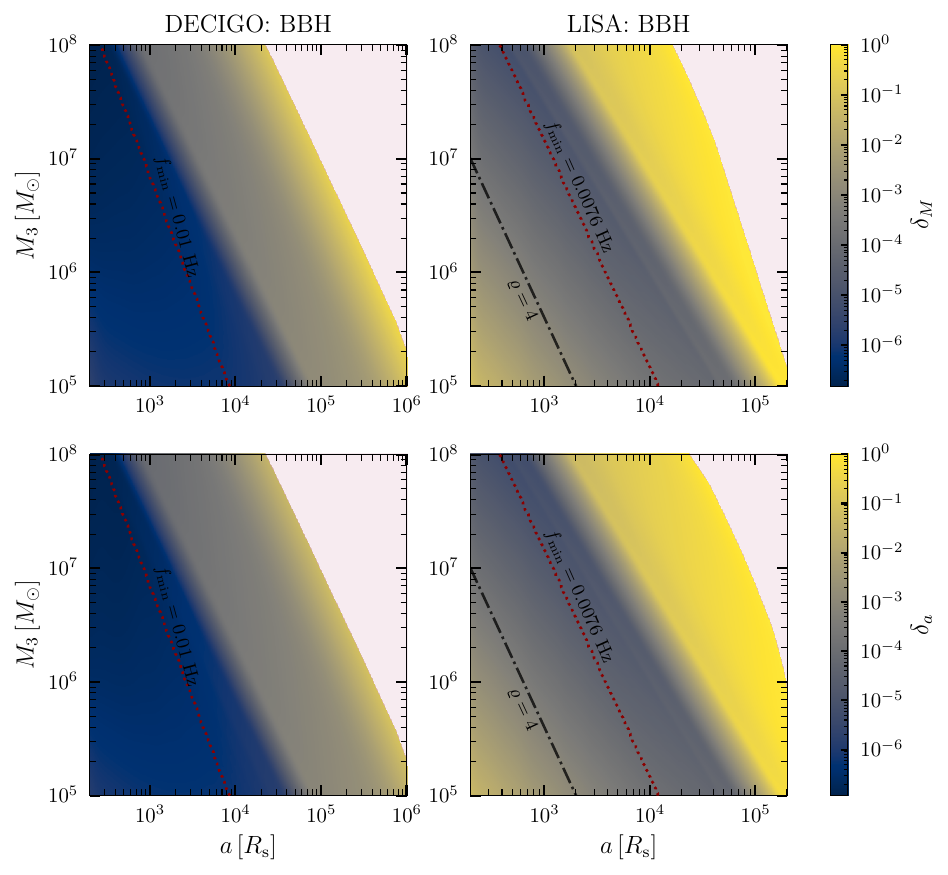}
    \caption{\textbf{SMBH:} {\it Left two panels} show the relative errors in the measurement of $M_3$ ({\it top panels}) and $a$ ({\it bottom panels}) over a grid of $M_3$ and $a$ for the A+: NSBH and ET: BBH2 cases mentioned in Table~\ref{tab: tab_sys} in COO scenario, while the {\it right two panels} show the same for DECIGO: BBH and LISA: BBH cases. Unlike other cases, here we have fixed $\theta_{\rm c} = 0.45$, i.e., $\cos \theta_{\rm c} \approx 0.9$ to compare our results against Figure 2 of~\cite{Tiwari:2024pvb} and update the same by filling in the parameter space (bottom left portion of the parameter space) where $\vert \Gamma_n t_{\rm obs} \vert \ll 1$ was not satisfied. The patches on the upper right and the dotted lines have the same meaning as in Figure~\ref{fig: Ap_ET_BNS_BBH_SBH_IMBH_CO}. The dashed-dotted lines in the rightmost panels represent the contours of a constant SNR and demarcate the region (lower left to the line) where the CBC is not detectable because the SNR falls below 4.}
    \label{fig: Ap_ET_DECIGO_LISA_CO_ME}
\end{figure*}

Following section~\ref{sec: constraints}, we estimate the constraints on the mass of the third body in the vicinity of the CBC and outer orbit parameters. We consider circular as well as eccentric outer orbits in several systems and detector configurations mentioned in Table~\ref{tab: tab_sys}. We consider a range of tertiary masses, $M_3$, spanning stellar-mass BHs (SBHs), intermediate-mass BHs (IMBHs), and supermassive BHs (SMBHs). Specifically, we consider masses in the range $1 - 10^8\,M_{\odot}$. Here, we would like to clarify that an object having mass in the range $1-5\,M_{\odot}$ is typically assumed not to be a BH under standard astrophysical scenarios; therefore, together with the stability criteria discussed in section~\ref{subsec: stab}, one also needs to ensure that the CBC does not tidally disrupt this object. We have checked that, assuming $1-5\,M_{\odot}$ objects to be NSs, the critical radius for the disruption of the tertiary also falls below the critical radius for stability. Therefore, as discussed in section~\ref{subsec: stab}, we only demarcate the regions where the system is unstable against escape. 

Throughout this {paper}, we fix $\theta_{\rm c} = \vartheta_{\rm p} = 0.1$ radians and $e_{\rm out}= 0.5$ for eccentric outer orbits, while fix $\theta_{\rm c} = 0.1$ radians for circular orbits unless specified otherwise. In addition, we set a lower limit of $a$ in all analyses to ensure $z_{\rm L,0} \leq 0.05$ in COOs and $z_{\rm L,0} / \sqrt{1 - e_{\rm out}^2} \leq 0.05$ in EOOs. For $M_3 \gg M_{\rm S}$, this sets the lower limit of $a$ to $200 \, R_{\rm s}$, while for a tertiary of finite mass, this limit changes depending on the CBC masses but remains below $\sim 270 \, R_{\rm s}$ for the systems considered in this analysis. We define the relative error in the measurement of a parameter $X$ as $\delta_X \equiv \Delta X / X$; e.g., we denote the relative error in the measurement of $M_3$ by $\delta_M$. In the upcoming subsections, we discuss the results.

\begin{figure*}[!ht]
    \centering
    \includegraphics[width=0.495\linewidth]{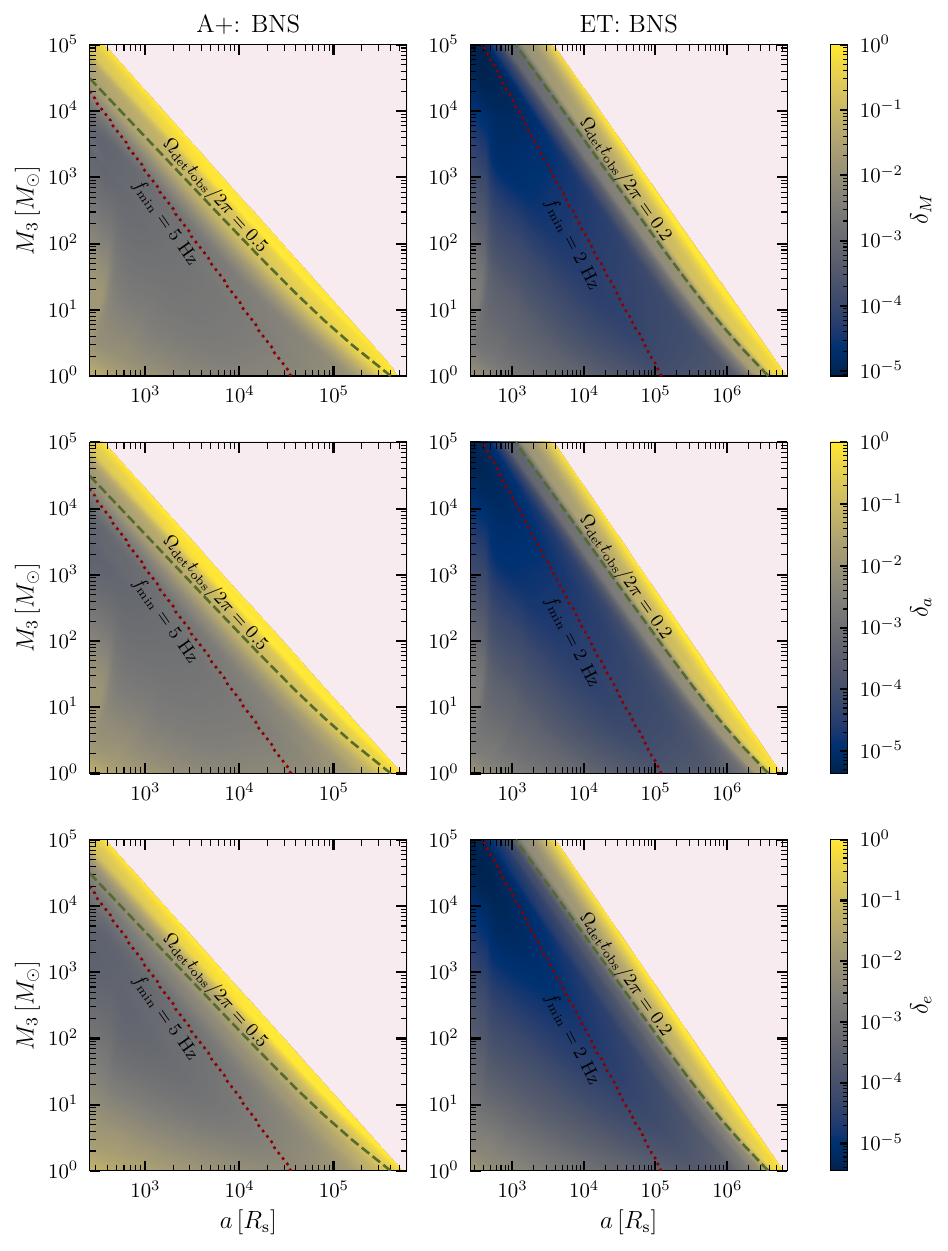}
    \includegraphics[width=0.495\linewidth]{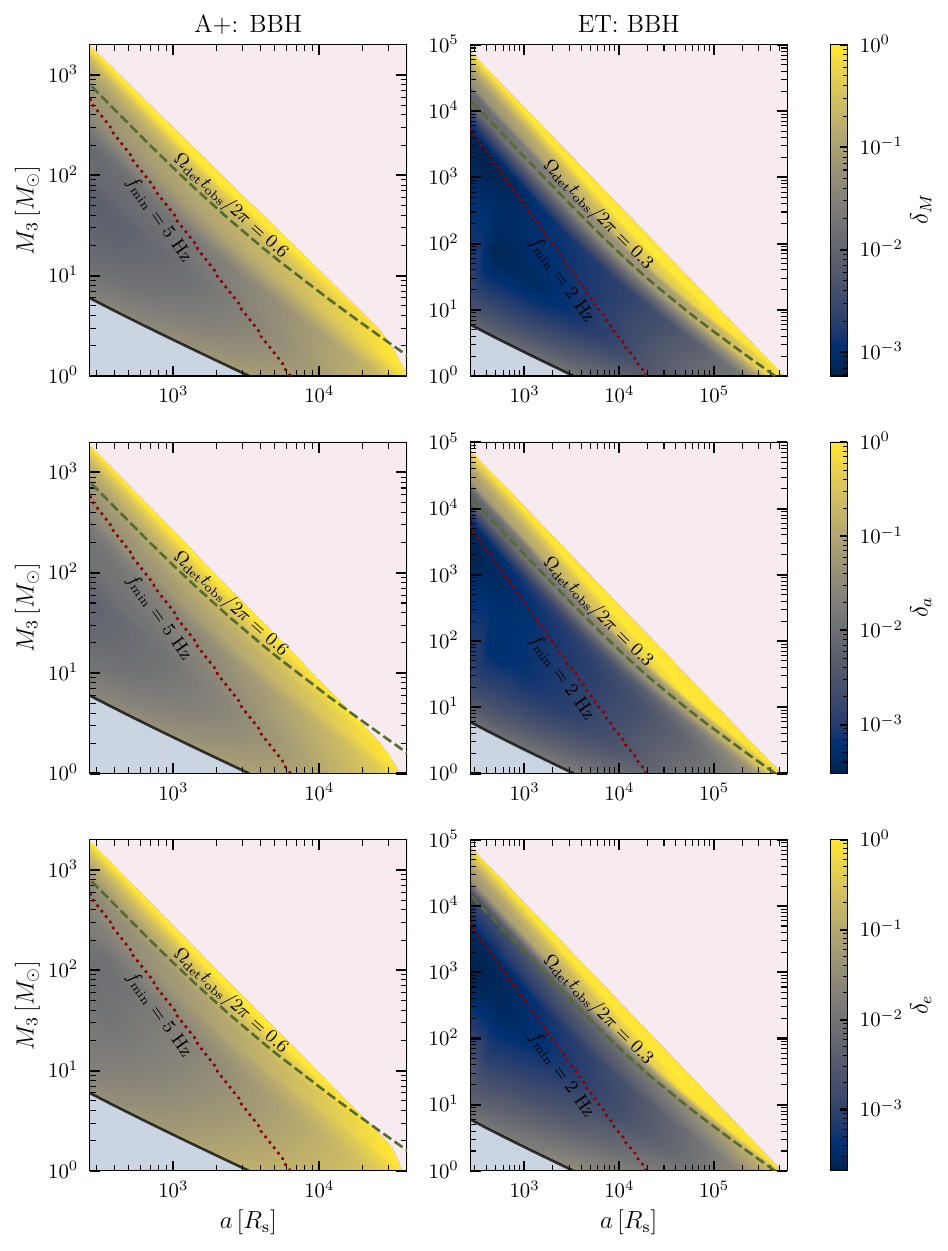}
    \caption{\textbf{SBH-IMBH:} {\it Left two panels} show the relative errors in the measurement of mass of the tertiary $M_3$ ({\it top panels}), semi-major axis of the outer orbit $a$ ({\it middle panels}), and eccentricity of the outer orbit $e_{\rm out}$ ({\it bottom panels}) over a grid of $M_3$ and $a$ for the A+: BNS and ET: BNS cases mentioned in Table~\ref{tab: tab_sys} in EOO scenario, while the {\it right two panels} show the same for A+: BBH and ET: BBH cases. The patches on the upper right and bottom left, and the dashed and dotted lines have the same meaning as in Figure~\ref{fig: Ap_ET_BNS_BBH_SBH_IMBH_CO}.}
    \label{fig: Ap_ET_BNS_BBH_SBH_IMBH_EO}
\end{figure*}

\subsection{Circular outer orbits}\label{subsec: circ_res}
Figure~\ref{fig: Ap_ET_BNS_BBH_SBH_IMBH_CO} shows the relative errors in the measurement of $M_3$ and $a$ over a grid of $M_3$ and $a$ for A+: BNS, ET: BNS, A+: BBH, and ET: BBH cases for SBH and IMBH-like tertiary masses in the COO scenarios. We observe that both $\delta_M$ and $\delta_a$, as well as $\delta_z$ and $\delta_{\Omega}$ (see Figure~\ref{fig: Ap_ET_BNS_BBH_SBH_IMBH_zo_CO}),  follow a diagonal line like pattern in $M_3 - a$ grid in log-log space. This is because in the $\Omega_{\rm det} t_{\rm obs} / 2 \pi \gg 1$ region of the parameter space, from Equation~\eqref{eq: ph_cor_losv_circ}, the magnitude of the phase correction varies as $z_{\rm L,0} / \Omega_{\rm det} \propto M_3^2 a / (M_3 + M_{\rm S})$ for tertiaries of finite mass and as $z_{\rm L,0} / \Omega_{\rm det} \propto M_3 a$ for tertiaries of mass $M_3 \gg M_{\rm S}$, while in the $\Omega_{\rm det} t_{\rm obs} / 2 \pi \ll 1$ region of the parameter space, the magnitude of the phase correction varies as $z_{\rm L,0} \Omega_{\rm det} \propto 1/(M_3 a^2)$ at the lowest order for tertiaries of all masses. These, in combination, lead to diagonal line-like patterns in log-log space. In addition, we see that $a$ is more precisely measured than $M_3$ and up to a larger distance, which is because it is mostly determined by the precision in the measurement of $z_{\rm L,0}$, while the precision in the measurement of $M_3$ depends on $\Omega_{\rm det}$ as well, especially for the massive IMBHs where Equation~\eqref{eq: zl0_fin} becomes equivalent to Equation~\eqref{eq: zl0_inf}. 

In Figure~\ref{fig: Ap_ET_BNS_BBH_SBH_IMBH_zo_CO} of the Appendix~\ref{app: add_fig}, we show the corresponding relative errors in the measurement of $z_{\rm L,0}$ and $\Omega_{\rm det}$. It explicitly shows that $z_{\rm L,0}$ is a more precisely measured quantity than $\Omega_{\rm det}$ and, as a byproduct, so is $a$. Note, however, that in the parameter space where only $z_{\rm L,0}$ and, as a result, $a$ are measurable, the measurability of $a$ does not give us any meaningful information regarding the tertiary because we can not identify what the tertiary is without its mass. In fact, in Appendix~\ref{app: add_fig_wav}, we show that the measurability of only $z_{\rm L,0}$ does not give us any additional information because it will become degenerate with $M$. Therefore, we say the tertiary is detectable only when we can constrain $M_3$ and $a$ both. We also see a slope change for the $\delta_a = 1$ boundary, which is a numerical artifact because we are in the $\Omega_{\rm det} t_{\rm obs} / 2 \pi \ll 1$ region --- see $\Omega_{\rm det} t_{\rm obs} / 2 \pi$ contours demarcating the same --- and as we go deeper in $\Omega_{\rm det} t_{\rm obs} / 2 \pi \ll 1$ region, the Fisher matrix in $z_{\rm L,0}$ and $\ln \Omega_{\rm det}$ becomes inefficient. A more convenient approach in this regime would be to parameterize the kinematics of the CBC’s CoM in terms of LOSA and its higher-order time derivatives \cite{Tiwari:2024pvb}. 

We find that a $1\, M _ {\odot}$ object in the vicinity of a $1.6-1.3 \, M_{\odot}$ BNS at $100 \, \rm Mpc$ can be detected up to $\sim 3 \times 10^5 \, R_{\rm s} \, (0.006 \, \rm AU)$ with A+ and up to $\sim 10^7 \, R_{\rm s} \, (0.2 \, \rm AU)$ with ET, and a $10^5 \, M_{\odot}$ SMBH in the vicinity of the same can be detected up to $\sim 800 \, R_{\rm s} \, (1.58 \, \rm AU)$ with A+ and up to $\sim 2 \times 10^4 \, R_{\rm s} \, (39.5 \, \rm AU)$ with ET. For the third body in the vicinity of a $10-10 \, M_{\odot}$ BBH at $500 \, \rm Mpc$ in A+ and $1 \, \rm Gpc$ in ET, we find that a $1 \, M_{\odot}$ object can be detected up to $\sim 4 \times 10^4 \, R_{\rm s} \, ( 8 \times 10^{-4} \, \rm AU)$ with A+ and up to $\sim 4 \times 10^5 \, R_{\rm s} \, (8 \times 10^{-3} \, \rm AU)$ with ET; a $10^3 \, M_{\odot}$ IMBH can be detected up to $700 \, R_{\rm s} \, ( 0.014 \, \rm AU)$ with A+ and $10^4 \, R_{\rm s} \, ( 0.2 \, \rm AU)$ with ET; and a $10^5 \, M_{\odot}$ IMBH can be detected up to $\sim 700 \, R_{\rm s} \, (1.38 \, \rm AU)$ with ET.

Figure~\ref{fig: Ap_ET_bns_SMBH} shows the relative errors in the measurement of $M_3$ and $a$ (see Figure~\ref{fig: Ap_ET_bns_SMBH_zo_CO} of Appendix~\ref{app: add_fig} for the constraints on $z_{\rm L,0}$ and $\Omega_{\rm det}$) over a grid of $M_3$ and $a$ for A+: BNS and ET: BNS cases for SMBH like tertiary masses in the COO scenarios. We find that a $4 \times 10^5 \, M_{\odot}$ SMBH in the vicinity of $1.6-1.3 \, M_{\odot}$ BNS at $100 \, \rm Mpc$ can be detected up to $\sim 400 \, R_{\rm s} \, (3.16 \, \rm AU)$ with A+ and up to $\sim 8 \times 10^3 \, R_{\rm s} \, (63.17 \, \rm AU)$ with ET. In addition, a $10^7 \, M_{\odot}$ and $10^8 \, M_{\odot}$ SMBHs in the vicinity of the same can be detected up to $\sim 10^3 \, R_{\rm s} \, (197.41 \, \rm AU)$ and up to $\sim 300 \, R_{\rm s} \, (592.24 \, \rm AU)$, respectively, with ET.

\begin{figure*}[!ht]
    \centering
    \includegraphics[width=0.8\linewidth]{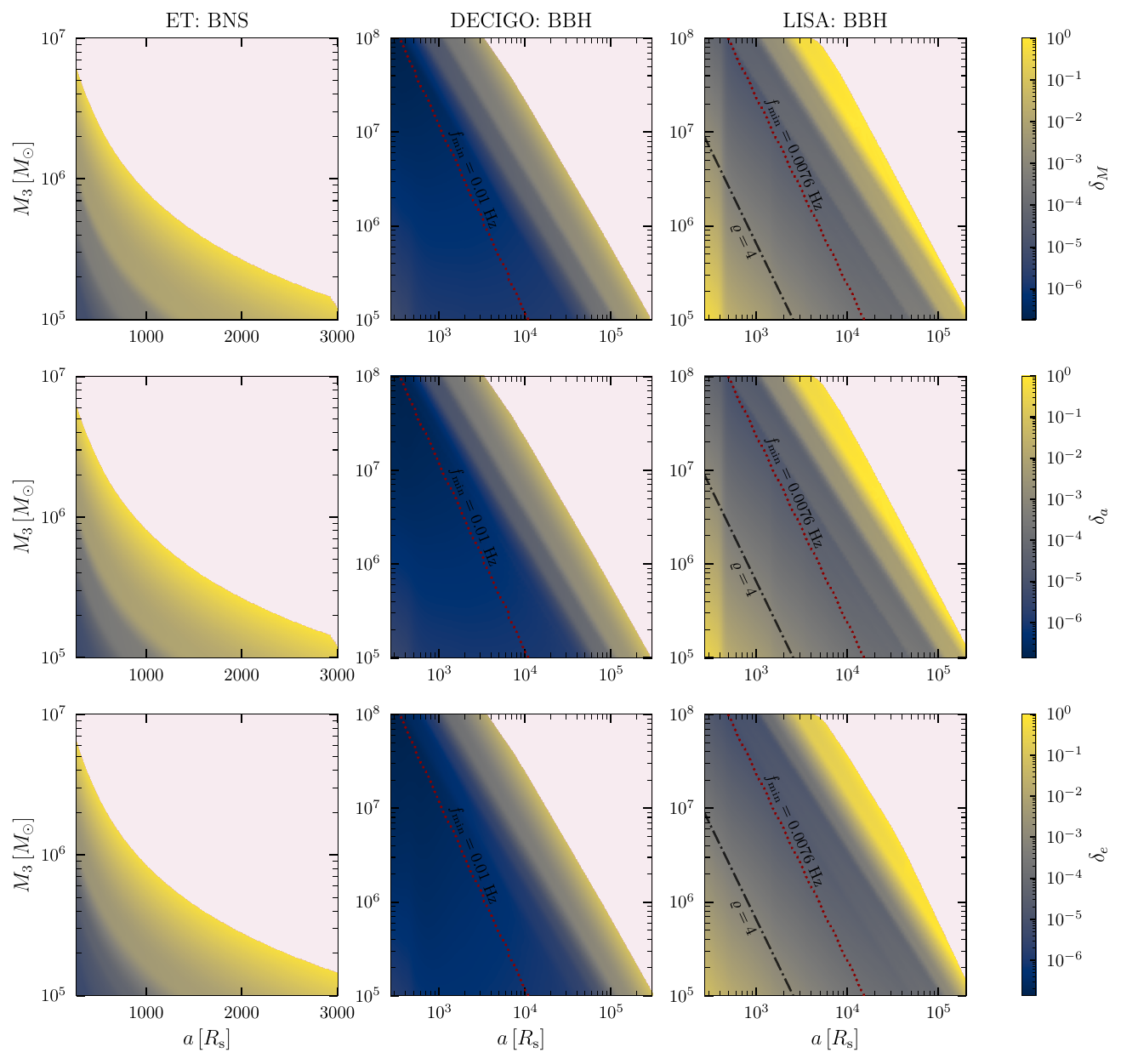}
    \caption{\textbf{SMBH:} {\it left panel} show the relative errors in the measurement of $M_3$ ({\it top panels}), $a$ ({\it middle panels}), and $e_{\rm out}$ ({\it bottom panels}) over a grid of $M_3$ and $a$ for the ET: BNS case mentioned in Table~\ref{tab: tab_sys} in EOO scenario, the {\it middle panel from left} shows the same for the DECIGO: BBH case, while {\it right panel} shows the same for the LISA: BBH case. The patches on the upper right and the dotted lines have the same meaning as in Figure~\ref{fig: Ap_ET_BNS_BBH_SBH_IMBH_CO}, while the dashed-dotted lines in the rightmost panels have the same meaning as in Figure~\ref{fig: Ap_ET_DECIGO_LISA_CO_ME}.}
    \label{fig: ET_DECIGO_LISA_SMBH}
\end{figure*}

Figure~\ref{fig: Ap_ET_DECIGO_LISA_CO_ME} shows the relative errors in the measurement of $M_3$ and $a$ (see Figure~\ref{fig: Ap_ET_BNS_BBH_SBH_IMBH_zo_CO} of Appendix~\ref{app: add_fig} for the constraints on $z_{\rm L,0}$ and $\Omega_{\rm det}$) over a grid of $M_3$ and $a$ for A+: NSBH, ET: BBH2, DECIGO: BBH, and LISA: BBH cases for SMBH like tertiary masses in the COO scenarios for $\theta_{\rm c} = 0.45$, i.e., $\cos \theta_{\rm c} \approx 0.9$. We chose a different value of $\theta_{\rm c}$ in this case to compare our results with Figure 2 of~\cite{Tiwari:2024pvb} and update the same by filling in the parameter space --- bottom left portions in the Figure 2 of~\cite{Tiwari:2024pvb}, where $\vert \Gamma_n t_{\rm obs} \vert \ll 1$ was not satisfied, $\Gamma_n$ being the $n^{th}$ time derivative of the LOSV. We also expand the parameter space down to $10^5 \, M_{\odot}$. We find improvements in all cases in comparison to Figure 2 of~\cite{Tiwari:2024pvb}, especially in the A+: NSBH and ET: BBH2 cases. Specifically, we find that a $10^5 \, M_{\odot}$ SMBH in the vicinity of a $5-1.4 \, M_{\odot}$ NSBH at 100 Mpc in A+ and a $30-30 \, M_{\odot}$ BBH at 100 Mpc in ET can be detected up to $\sim 400 \, R_{\rm s} \, (0.79 \, \rm AU)$.
We also find that a $10^5 \, M_{\odot}$ SMBH in the vicinity of a $100-100 \, M_{\odot}$ BBH at 1 Gpc in DECIGO and LISA can be detected up to $> 10^6 \, R_{\rm s} \, (0.01 \, \rm pc)$ and $\sim 2 \times 10^5 \, R_{\rm s} \, (0.002 \, \rm pc)$, respectively, while a $10^8 \, M_{\odot}$ SMBH in the vicinity of the same can be detected up to $\gtrsim 2 \times 10^4 \, R_{\rm s} \, (0.19 \, \rm pc)$ with both of them.

Note that for DECIGO: BBH and LISA: BBH cases, as we enter in the deep $\Omega_{\rm det} t_{\rm obs} / 2 \pi \ll 1$ region such that only the leading order terms of the sine and cosine series expansion in Equation~\eqref{eq: ph_cor_losv_circ} become dominant, the Fisher matrix inversion in terms of $z_{\rm L,0}$ and $\ln \Omega_{\rm det}$ becomes inefficient because the matrix becomes roughly singular and one would be required to parametrize the kinematics of the CBC's CoM in terms of LOSA and its higher order time derivatives \cite{Tiwari:2024pvb}.

\subsection{Eccentric outer orbits}\label{subsec: ecc_res}
Figure~\ref{fig: Ap_ET_BNS_BBH_SBH_IMBH_EO} shows the relative errors in the measurement of $M_3$, $a$, and $e_{\rm out}$  (see Figure~\ref{fig: Ap_ET_BNS_BBH_SBH_IMBH_zo_EO} of Appendix~\ref{app: add_fig} for the constraints on $z_{\rm L,0}$ and $\Omega_{\rm det}$) over a grid of $M_3$ and $a$ for A+: BNS, ET: BNS, A+: BBH, and ET: BBH cases for SBH and IMBH like tertiary masses in the EOO scenarios.  We observe that all of $\delta_M$, $\delta_a$, and $\delta_e$, as well as $\delta_z$ and $\delta_{\Omega}$ (see Figure~\ref{fig: Ap_ET_BNS_BBH_SBH_IMBH_zo_EO}), follow the similar pattern as the COO cases, except for the slope change in $\delta_a = 1$ boundaries in A+: BBH and ET: BBH cases because $\delta_a$ has become greater than 1 in those regions, which is because we have increased the number of parameters in the Fisher matrix and hence increased errors in the measurement of all parameters. Due to this same reason, unlike the COO cases (see Figure~\ref{fig: Ap_ET_BNS_BBH_SBH_IMBH_CO}), $M_3$, $a$, and $e_{\rm out}$ share the same boundaries of $\delta_M = 1$, $\delta_a = 1$, and $\delta_e = 1$. 

We find that a $1\, M_{\odot}$ object in the vicinity of a $1.6-1.3 \, M_{\odot}$ BNS at $100 \, \rm Mpc$ can be detected in an EOO of eccentricity 0.5 and a semi-major axis of up to $\sim 4 \times 10^5 \, R_{\rm s} \, (0.008 \, \rm AU)$ with A+ and up to $\sim 6 \times 10^6 \, R_{\rm s} \, (0.12 \, \rm AU)$ with ET, and a $10^5 \, M_{\odot}$ SMBH in the vicinity of the same can be detected up to $\sim 400 \, R_{\rm s} \, (0.79 \, \rm AU)$ with A+ and up to $\sim 4 \times 10^3 \, R_{\rm s} \, (7.90 \, \rm AU)$ with ET. For the tertiary in the vicinity of a $10-10 \, M_{\odot}$ BBH at $500 \, \rm Mpc$ in A+ and $1 \, \rm Gpc$ in ET, we find that a $1 \, M_{\odot}$ object can be detected in an EOO of eccentricity 0.5 and a semi-major axis of up to $\sim 4 \times 10^4 \, R_{\rm s} \, ( 8 \times 10^{-4} \, \rm AU)$ with A+ and up to $\sim 5 \times 10^5 \, R_{\rm s} \, (10^{-2} \, \rm AU)$ with ET; a $10^2 \, M_{\odot}$ SBH can be detected up to $2 \times 10^3 \, R_{\rm s} \, (4 \times 10^{-3} \, \rm AU)$ with A+ and $2 \times 10^4 \, R_{\rm s} \, ( 0.04 \, \rm AU)$ with ET; and a $10^4 \, M_{\odot}$ IMBH can be detected in an EOO of eccentricity 0.5 and a semi-major axis of up to $\sim 10^3 \, R_{\rm s} \, (0.20 \, \rm AU)$ with ET.

Figure~\ref{fig: ET_DECIGO_LISA_SMBH} shows the relative errors in the measurement of $M_3$, $a$, and $e_{\rm out}$ (see Figure~\ref{fig: ET_DECIGO_LISA_SMBH_zo_EO} of Appendix~\ref{app: add_fig} for the constraints on $z_{\rm L,0}$ and $\Omega_{\rm det}$) over a grid of $M_3$ and $a$ for ET: BNS, DECIGO: BBH, and LISA: BBH cases for SMBH like tertiary masses in the EOO scenarios. We find that $10^5 \, M_{\odot}$ and $10^6 \, M_{\odot}$ SMBHs in the vicinity of a $1.6-1.3 \, M_{\odot}$ BNS at $100 \, \rm Mpc$ can be detected in an EOO of eccentricity 0.5 and a semi-major axis of up to $\sim 3 \times 10^3 \, R_{\rm s} \, (5.92 \, \rm AU)$ and $\sim 800 \, R_{\rm s} \, (15.79 \, \rm AU)$, respectively, with ET. We further find that $10^5 \, M_{\odot}$ and $10^8 \, M_{\odot}$ SMBHs in the vicinity of a $100-100 \, M_{\odot}$ BBH at 1 Gpc can be detected in an EOO of eccentricity 0.5 and a semi-major axis of up to $\sim 3 \times 10^5 \, R_{\rm s} \, (592.24 \, \rm AU)$ and $\sim 3 \times 10^3 \, R_{\rm s} \, (0.029 \, \rm pc)$, respectively, with DECIGO and $\sim 2 \times 10^5 \, R_{\rm s} \, (394.83 \, \rm AU)$ and $\sim 4 \times 10^3 \, R_{\rm s} \, (0.038 \, \rm pc)$, respectively, with LISA.

\section{Discussion}
Extracting information encoded in GW signals about merger environments is important to understand the astrophysical origins of CBCs. In Ref.~\cite{Tiwari:2024pvb}, we showed that by studying the kinematics of a CBC's CoM through the imprints of its LOSV varying quadratically, cubically, or higher orders with time onto the GW waveform, precise information about the merger environments --- such as the mass of the host potential, location of the CBC within the potential, and the steepness of the potential profile --- can be extracted on a {\it single-event basis}. In the context of circular and eccentric outer orbits, these approximations to the LOSV of a CBC's CoM are equivalent to observing only a small segment of the outer orbit when the outer orbital period is very large compared to the observation time of a CBC. 

In this {\it paper}, we derived the leading-order phase and amplitude corrections to the $(\ell, \, m) = (2, \, 2)$ mode of the GW waveform due to the LOSV of the CBC's CoM in the cases of circular and eccentric outer orbits. These corrections are valid for all outer orbital periods in the limits $z_{\rm L,0} \ll 1$ and $z_{\rm L,0} / \sqrt{1 - e_{\rm out}^2} \ll 1$ for COOs and EOOs, respectively, and can be used for any frequency range. Contrast this with ~\cite{Tiwari:2024pvb}, where $f_{\rm max}$ needs to be either $f_{\rm lso}$ or a value closer to it such that the time to coalescence at $f_{\rm max} \ll$ observation time. This is because LOSV, in Ref.~\cite{Tiwari:2024pvb}, has been Taylor expanded about $t_{\rm c}$. 

Though amplitude corrections for higher modes still need to be computed separately, the phase corrections can be applied to any higher mode by using the transformation $\Delta \Psi (f) \to (m / 2) \Delta \Psi (2f/m)$. We show that the morphology of the GW waveform of a CBC subjected to a periodic motion is quite different from that of a static CBC, especially when the outer orbital period is smaller than the observation time. Specifically, we find that the time-domain unperturbed and perturbed waveforms of a CBC subjected to a periodic motion go in and out of phase repeatedly when the outer orbital period is smaller than the signal duration.

We considered several system configurations of BBHs, BNSs, and NSBHs in the sensitivity bands of next-generation ground- and space-based detectors, namely A+, ET, DECIGO, and LISA, in the circular and eccentric outer orbits, and calculated the relative errors in the measurement of the mass of the third body\footnote{In general, the third body can also be an exoplanet. We refer the reader to~\cite{Tiwari:2026cnb}, in which we specifically consider the circum-CBC exoplanets and show that these could be detectable even with the next-generation ground-based detectors such as ET.}, radius/semi-major axis of the outer orbit, and eccentricity of the outer orbit. We showed that precise information about the mass of the third body in the vicinity of a CBC, and its distance from the CBC, can be obtained by extracting parameters related to the LOSV of the CBC's CoM, on a {\it single event basis}. We also updated Figure 2 of~\cite{Tiwari:2024pvb} by expanding the parameter space.

In this {\it work}, while conducting the Fisher matrix analysis, we have inherently assumed that we know the true model, which would not be the case in general while performing Bayesian parameter estimation on real data. However, as shown in Ref.~\cite{Tiwari:2024pvb}, a simple Bayesian model selection between two models will be able to pick the correct model. 

We expect our work to have significant applications to binary-single and binary-binary encounters leading to hierarchical triples in dynamical environments such as globular clusters~\cite{banerjee2010, banerjee2018, Chatterjee2017, Chatterjee2017a, fragione2018, rodriguez2018, dicarlo2020, kremer2020, mapelli2021, trani2021, FragioneRasio2023} and nuclear star clusters~\cite{antonini2012, petrovich2017, grishin2018, hoang2018, fragione2020}. It has also been suggested that BBHs, BNSs, and NSBHs can form in AGN migration traps~\cite{McKernan:2020lgr, Tagawa:2019osr}, while~\cite{Tagawa:2026afw} suggests mergers can also form outside the migration traps. This work can therefore be used to test these models because a small fraction of them having smaller masses (see~\cite{Tagawa:2026afw} for the distributions of CBC masses and their merger distances from the SMBH for a Milky-Way-like Galaxy) could have a detectable imprint of LOSV in ET (see Figures~\ref{fig: Ap_ET_bns_SMBH} and~\ref{fig: Ap_ET_DECIGO_LISA_CO_ME}). 

Finally, we have considered only one set of fiducial values of $e_{\rm out}$, $\theta_{\rm c}$, and $\vartheta_{\rm p}$. The formalism, however, is valid for any set of values of these parameters except for $e_{\rm out} \gtrsim 0.66$; therefore, it will be interesting to see how varying these parameters would affect the constraints on the LOSV parameters and, in turn, on $M_3$ and $a$. In addition, it is worthwhile to study how incorporating tidal dephasing in the unperturbed (without LOSV) waveform affects the measurability of the LOSV parameters. Since the corrections in this {\it paper} have been derived only at Newtonian order, we plan to follow it up by calculating corrections for higher PN orders. We also intend to extend the framework to EOOs with $e_{\rm out} \gtrsim 0.66$.

\begin{acknowledgments}
    We thank Lalit Pathak for the LVK’s internal Publications and Presentations review of this work and for his careful reading and suggestions. We also thank Nathan Johnson-McDaniel for useful discussions and suggestions. SJK acknowledges support from ANRF/SERB Grants SRG/2023/000419 and MTR/2023/000086.
\end{acknowledgments}

\vspace{5mm}

\textit{Software}: \texttt{NumPy} \cite{vanderWalt:2011bqk}, \texttt{SciPy} \cite{Virtanen:2019joe}, \texttt{astropy} \cite{2013A&A...558A..33A, 2018AJ....156..123A}, \texttt{Matplotlib} \cite{Hunter:2007}, \texttt{jupyter} \cite{jupyter}, \texttt{LALSuite}~\cite{2020ascl.soft12021L}, \texttt{Bilby}~\cite{bilby_paper}, \texttt{PyCBC}~\cite{alex_nitz_2024_10473621}.


\bibliography{references}

\onecolumngrid

\appendix

\section{Eccentric Outer Orbits}\label{app: eoo_}

\subsection{$\cos \vartheta$ and $\sin \vartheta$ Expansions}\label{app: ecc_har}
The expansions of $\cos \vartheta$ and $\sin \vartheta$ in $e_{\rm out}$ and $\zeta$ used the Section~\ref{subsec: ecc_der} are given by
\begin{multline}
    \label{eq: cos_varth}
    \cos \vartheta =  e_{\rm out}^4 \left(\frac{25}{192}\cos \zeta - \frac{225}{128} \cos 3 \zeta  +\frac{625}{384} \cos 5 \zeta \right) + \frac{4}{3} e_{\rm out}^3 (\cos 4 \zeta - \cos 2 \zeta ) \\ + \frac{9}{8} e_{\rm out}^2 (\cos 3 \zeta - \cos \zeta ) + e_{\rm out} (\cos 2 \zeta - 1) + \cos \zeta
\end{multline}
\begin{multline}
    \label{eq: sin_varth}
    \sin \vartheta =  e_{\rm out}^4 \left(\frac{17}{192}\sin \zeta -\frac{207}{128} \sin 3 \zeta  + \frac{625}{384} \sin 5 \zeta \right) + e_{\rm out}^3 \left(\frac{4}{3} \sin 4 \zeta -\frac{7}{6} \sin 2 \zeta \right)\\  + e_{\rm out}^2 \left(\frac{9}{8} \sin 3 \zeta - \frac{7 }{8} \sin \zeta \right) + e_{\rm out} \sin 2 \zeta +\sin \zeta 
\end{multline}

\subsection{Time and Orbital Phase}\label{app: time_ph_eoo}
The time and orbital phase for the EOO case are given by

\begin{multline}
    \label{eq: neg_coal_t_ecc}
    (t - t_c)_{\rm EL} = -\frac{5 M}{256 \eta  v^8} \Biggl[ 1 + \frac{z_{\rm L,0}}{\sqrt{1-e^2_{\rm out}}} \Biggl[\frac{v^8}{\xi} \left( \sin \left(\frac{\xi }{v^8} - \theta_{\rm c}-\vartheta_{\rm p} \right) - \sin \left(\frac{\xi }{v_{\rm lso}^8} - \theta_{\rm c}-\vartheta_{\rm p} \right) \right) - \frac{8}{3} \cos \left(\frac{\xi }{v^8} - \theta_{\rm c}-\vartheta_{\rm p} \right) \\
    + \Biggl\{\frac{v^8 }{2 \xi } \left(\sin \left(\frac{2 \xi }{v^8} - 2 \theta_{\rm c}-\vartheta_{\rm p} \right) - \sin \left(\frac{2 \xi }{v_{\rm lso}^8} - 2 \theta_{\rm c}-\vartheta_{\rm p} \right)  \right) -\frac{8}{3} \cos \left(\frac{2 \xi }{v^8} - 2 \theta_{\rm c}-\vartheta_{\rm p} \right) \Biggr\} e_{\rm out} 
    + \Biggl\{\frac{3 v^8}{8 \xi } \Biggl( \sin \left(\frac{3 \xi }{v^8} - 3 \theta_{\rm c}-\vartheta_{\rm p} \right)  \\ - \sin \left(\frac{3 \xi }{v_{\rm lso}^8} - 3 \theta_{\rm c}-\vartheta_{\rm p} \right) \Biggr) -\frac{v^8 }{\xi } \left( \sin \left(\frac{\xi }{v^8} - \theta_{\rm c}-\vartheta_{\rm p} \right) - \sin \left(\frac{\xi }{v_{\rm lso}^8} - \theta_{\rm c}-\vartheta_{\rm p} \right) \right)  - \frac{v^8 }{8 \xi } \Biggl( \sin \left(\frac{\xi }{v^8} - \theta_{\rm c}+\vartheta_{\rm p} \right) \\ - \sin \left(\frac{\xi }{v_{\rm lso}^8} - \theta_{\rm c}+\vartheta_{\rm p} \right) \Biggr)  +\frac{8}{3} \cos \left(\frac{\xi }{v^8} - \theta_{\rm c}-\vartheta_{\rm p} \right)-3 \cos \left(\frac{3 \xi }{v^8} - 3 \theta_{\rm c}-\vartheta_{\rm p} \right)+\frac{1}{3} \cos \left(\frac{\xi }{v^8} - \theta_{\rm c}+\vartheta_{\rm p} \right)  \Biggr\} e^2_{\rm out} 
    \\ + \Biggl\{ \frac{v^8 }{3 \xi } \left( \sin \left(\frac{4 \xi }{v^8} - 4 \theta_{\rm c}-\vartheta_{\rm p} \right) - \sin \left(\frac{4 \xi }{v_{\rm lso}^8} - 4 \theta_{\rm c}-\vartheta_{\rm p} \right) \right)  - \frac{5 v^8 }{8 \xi } \left( \sin \left(\frac{2 \xi }{v^8} - 2 \theta_{\rm c}-\vartheta_{\rm p} \right) - \sin \left(\frac{2 \xi }{v_{\rm lso}^8} - 2 \theta_{\rm c}-\vartheta_{\rm p} \right) \right)  \\ - \frac{v^8 }{24 \xi } \left( \sin \left(\frac{2 \xi }{v^8} - 2 \theta_{\rm c}+\vartheta_{\rm p} \right) - \sin \left(\frac{2 \xi }{v_{\rm lso}^8} - 2 \theta_{\rm c}+\vartheta_{\rm p} \right) \right) +\frac{10}{3} \cos \left(\frac{2 \xi }{v^8} - 2 \theta_{\rm c}-\vartheta_{\rm p} \right)-\frac{32}{9} \cos \left(\frac{4 \xi }{v^8} - 4 \theta_{\rm c}-\vartheta_{\rm p} \right) \\ +\frac{2}{9} \cos \left(\frac{2 \xi }{v^8} - 2 \theta_{\rm c}+\vartheta_{\rm p} \right) \Biggr\} e^3_{\rm out} 
    + \Biggl\{ \frac{7 v^8 }{64 \xi } \left( \sin \left(\frac{\xi }{v^8} - \theta_{\rm c}-\vartheta_{\rm p} \right) - \sin \left(\frac{\xi }{v_{\rm lso}^8} - \theta_{\rm c}-\vartheta_{\rm p} \right) \right)  + \frac{125 v^8 }{384 \xi } \Biggl( \sin \left(\frac{5 \xi }{v^8} - 5 \theta_{\rm c}-\vartheta_{\rm p} \right) \\ - \sin \left(\frac{5 \xi }{v_{\rm lso}^8} - 5 \theta_{\rm c}-\vartheta_{\rm p} \right) \Biggr) +\frac{v^8 }{48 \xi } \left( \sin \left(\frac{\xi }{v^8} - \theta_{\rm c}+\vartheta_{\rm p} \right) - \sin \left(\frac{\xi }{v_{\rm lso}^8} - \theta_{\rm c}+\vartheta_{\rm p} \right) \right)  - \frac{9 v^8 }{16 \xi } \Biggl( \sin \left(\frac{3 \xi }{v^8} - 3 \theta_{\rm c}-\vartheta_{\rm p} \right) \\ - \sin \left(\frac{3 \xi }{v_{\rm lso}^8} - 3 \theta_{\rm c}-\vartheta_{\rm p} \right) \Biggr) -\frac{3 v^8 }{128 \xi } \left( \sin \left(\frac{3 \xi }{v^8} - 3 \theta_{\rm c}+\vartheta_{\rm p} \right) - \sin \left(\frac{3 \xi }{v_{\rm lso}^8} - 3 \theta_{\rm c}+\vartheta_{\rm p} \right) \right) -\frac{7}{24} \cos \left(\frac{\xi }{v^8} - \theta_{\rm c}-\vartheta_{\rm p} \right) \\ +\frac{9}{2} \cos \left(\frac{3 \xi }{v^8} - 3 \theta_{\rm c}-\vartheta_{\rm p} \right)-\frac{625}{144} \cos \left(\frac{5 \xi }{v^8} - 5 \theta_{\rm c}-\vartheta_{\rm p} \right)  -\frac{1}{18} \cos \left(\frac{\xi }{v^8} - \theta_{\rm c}+\vartheta_{\rm p} \right)+\frac{3}{16} \cos \left(\frac{3 \xi }{v^8} - 3 \theta_{\rm c}+\vartheta_{\rm p} \right)\Biggr\} e^4_{\rm out}  \Biggr] \Biggr]
\end{multline}
and
\begin{multline}
    \label{eq: phi_ecc}
    (\phi - \phi_{\rm c})_{\rm EL} = - \frac{1}{32 \eta  v^5 } \Biggl[1 - \frac{z_{\rm L,0}}{\sqrt{1-e^2_{\rm out}}} \Biggl[ \frac{5}{3} \cos \left(\frac{\xi }{v^8} - \theta_{\rm c}-\vartheta_{\rm p} \right) + \frac{5}{3} \cos \left(\frac{2 \xi }{v^8} - 2 \theta_{\rm c}-\vartheta_{\rm p} \right) e_{\rm out} 
    - \Biggl\{\frac{5}{3} \cos \left(\frac{\xi }{v^8} - \theta_{\rm c}-\vartheta_{\rm p} \right) \\ -\frac{15}{8} \cos \left(\frac{3 \xi }{v^8} - 3 \theta_{\rm c}-\vartheta_{\rm p} \right)+\frac{5}{24} \cos \left(\frac{\xi }{v^8} - \theta_{\rm c}+\vartheta_{\rm p} \right) \Biggr\} e^2_{\rm out} 
    - \Biggl\{ \frac{25}{12} \cos \left(\frac{2 \xi }{v^8} - 2 \theta_{\rm c}-\vartheta_{\rm p} \right)-\frac{20}{9} \cos \left(\frac{4 \xi }{v^8} - 4 \theta_{\rm c}-\vartheta_{\rm p} \right) \\ +\frac{5}{36} \cos \left(\frac{2 \xi }{v^8} - 2 \theta_{\rm c}+\vartheta_{\rm p} \right)  \Biggr\} e^3_{\rm out} 
    + \Biggl\{\frac{35}{192} \cos \left(\frac{\xi }{v^8} - \theta_{\rm c}-\vartheta_{\rm p} \right) - \frac{45}{16} \cos \left(\frac{3 \xi }{v^8} - 3 \theta_{\rm c}-\vartheta_{\rm p} \right) + \frac{5}{144} \cos \left(\frac{\xi }{v^8} - \theta_{\rm c}+\vartheta_{\rm p} \right) \\ - \frac{15}{128} \cos \left(\frac{3 \xi }{v^8} - 3 \theta_{\rm c}+\vartheta_{\rm p} \right) + \frac{3125}{1152} \cos \left(\frac{5 \xi }{v^8} - 5 \theta_{\rm c}-\vartheta_{\rm p} \right) \Biggr\} e^4_{\rm out}  \Biggr] \Biggr]\,,
\end{multline}
respectively.

\subsection{Phase and Amplitude Correction Coefficients}\label{app: ph_amp_corr_eoo}
The phase and amplitude correction coefficients ($P_n$ and $A_n$) of $e_{\rm out}^n$ are given by
\begin{equation}
    \label{eq: P0_ecc}
    P_0 = \sin \left(\frac{\xi}{v^8} - \theta_{\rm c} - \vartheta_{\rm p} \right) - \sin \left(\frac{\xi}{v_{\rm lso}^8} - \theta_{\rm c} - \vartheta_{\rm p} \right)
\end{equation}

\begin{equation}
    \label{eq: P1_ecc}
    P_1 = \frac{1}{2}\Biggl\{ \sin \left(\frac{2\xi}{v^8} - 2 \theta_{\rm c} - \vartheta_{\rm p} \right) - \sin \left(\frac{2\xi}{v_{\rm lso}^8} - 2 \theta_{\rm c} - \vartheta_{\rm p} \right)  \Biggr\}
\end{equation}

\begin{multline}
    \label{eq: P2_ecc}
    P_2 = - \left( \sin \left(\frac{\xi}{v^8} - \theta_{\rm c} - \vartheta_{\rm p} \right) - \sin \left(\frac{\xi}{v_{\rm lso}^8} - \theta_{\rm c} - \vartheta_{\rm p} \right) \right)
    + \frac{3}{8} \left( \sin \left(\frac{3\xi}{v^8} - 3 \theta_{\rm c} - \vartheta_{\rm p} \right) - \sin \left(\frac{3\xi}{v_{\rm lso}^8} - 3 \theta_{\rm c} - \vartheta_{\rm p} \right) \right) \\ - \frac{1}{8} \left( \sin \left(\frac{\xi}{v^8} - \theta_{\rm c} + \vartheta_{\rm p} \right) - \sin \left(\frac{\xi}{v_{\rm lso}^8} - \theta_{\rm c} + \vartheta_{\rm p} \right) \right)
\end{multline}

\begin{multline}
    \label{eq: P3_ecc}
    P_3 = - \frac{5}{8} \left(\sin \left(\frac{2\xi}{v^8} - 2 \theta_{\rm c} - \vartheta_{\rm p} \right) - \sin \left(\frac{2\xi}{v_{\rm lso}^8} - 2 \theta_{\rm c} - \vartheta_{\rm p} \right) \right) + \frac{1}{3} \left( \sin \left(\frac{4\xi}{v^8} - 4 \theta_{\rm c} - \vartheta_{\rm p} \right) -  \sin \left(\frac{4\xi}{v_{\rm lso}^8} - 4 \theta_{\rm c} - \vartheta_{\rm p} \right) \right) 
    \\ - \frac{1}{24} \left( \sin \left(\frac{2\xi}{v^8} - 2 \theta_{\rm c} + \vartheta_{\rm p} \right) - \sin \left(\frac{2\xi}{v_{\rm lso}^8} - 2 \theta_{\rm c} + \vartheta_{\rm p} \right) \right) 
\end{multline}

\begin{multline}
    \label{eq: P4_ecc}
    P_4 = - \frac{7}{64} \left( \sin \left(\frac{\xi}{v^8} - \theta_{\rm c} - \vartheta_{\rm p} \right) - \sin \left(\frac{\xi}{v_{\rm lso}^8} - \theta_{\rm c} - \vartheta_{\rm p} \right) \right) + \frac{9}{16} \Biggl( \sin \left(\frac{3\xi}{v^8} - 3 \theta_{\rm c} - \vartheta_{\rm p} \right) - \sin \left(\frac{3\xi}{v_{\rm lso}^8} - 3 \theta_{\rm c} - \vartheta_{\rm p} \right)\Biggr) 
    \\ - \frac{125}{384} \left( \sin \left(\frac{5\xi}{v^8} - 5 \theta_{\rm c} - \vartheta_{\rm p} \right) - \sin \left(\frac{5\xi}{v_{\rm lso}^8} - 5 \theta_{\rm c} - \vartheta_{\rm p} \right) \right) - \frac{1}{48} \Biggl(\sin \left(\frac{\xi}{v^8} - \theta_{\rm c} + \vartheta_{\rm p} \right) - \sin \left(\frac{\xi}{v_{\rm lso}^8} - \theta_{\rm c} + \vartheta_{\rm p} \right) \Biggr) \\ + \frac{3}{128} \left( \sin \left(\frac{3\xi}{v^8} - 3 \theta_{\rm c} + \vartheta_{\rm p} \right) - \sin \left(\frac{3\xi}{v_{\rm lso}^8} - 3 \theta_{\rm c} + \vartheta_{\rm p} \right)\right)
\end{multline}

\begin{equation}
    \label{eq: A0_ecc}
    A_0 = \frac{4}{3} \frac{\xi}{v^8} \sin \left(\frac{\xi}{v^8} - \theta_{\rm c} - \vartheta_{\rm p} \right) - \frac{1}{6} \cos \left(\frac{\xi}{v^8} - \theta_{\rm c} - \vartheta_{\rm p} \right)
\end{equation}

\begin{equation}
    \label{eq: A1_ecc}
    A_1 = \frac{8}{3} \frac{\xi}{v^8} \sin \left(\frac{2 \xi}{v^8} - 2 \theta_{\rm c} - \vartheta_{\rm p} \right) - \frac{1}{6} \cos \left(\frac{2 \xi}{v^8} - 2 \theta_{\rm c} - \vartheta_{\rm p} \right)
\end{equation}

\begin{multline}
    \label{eq: A2_ecc}
    A_2 = \frac{\xi}{v^8} \left\{ \frac{9}{2} \sin \left(\frac{3 \xi }{v^8} - 3 \theta _c-\theta _p \right) -\frac{4}{3} \sin \left(\frac{\xi }{v^8} - \theta _c -\theta _p \right) -\frac{1}{6} \sin \left(\frac{\xi }{v^8} -\theta _c+\theta _p \right) \right\} \\ +\frac{1}{6} \cos \left(\frac{\xi }{v^8} - \theta _c - \theta _p \right) + \frac{1}{48} \cos \left(\frac{\xi }{v^8} -\theta _c+\theta _p \right) -\frac{3}{16} \cos \left(\frac{3 \xi }{v^8} - 3 \theta _c-\theta _p \right)
\end{multline}

\begin{multline}
    \label{eq: A3_ecc}
    A_3 = \frac{\xi}{v^8} \left\{ \frac{64}{9} \sin \left(\frac{4 \xi }{v^8} - 4 \theta _c-\theta _p \right) -\frac{10}{3} \sin \left(\frac{2 \xi }{v^8} - 2 \theta _c-\theta _p \right) -\frac{2}{9} \sin \left(\frac{2 \xi }{v^8} - 2 \theta _c+\theta _p \right) \right\} \\ + \frac{5}{24} \cos \left(\frac{2 \xi }{v^8} - 2 \theta _c-\theta _p \right) +\frac{1}{72} \cos \left(\frac{2 \xi }{v^8} - 2 \theta _c+\theta _p \right) -\frac{2}{9} \cos \left(\frac{4 \xi }{v^8} - 4 \theta _c-\theta _p \right)   
\end{multline}

\begin{multline}
    \label{eq: A4_ecc}
    A_4 = \frac{\xi}{v^8} \Biggl\{ \frac{7}{48} \sin \left(\frac{\xi }{v^8} - \theta _c-\theta _p \right) +\frac{3125}{288} \sin \left(\frac{5 \xi }{v^8} - 5 \theta _c-\theta _p \right) + \frac{1}{36} \sin \left(\frac{\xi }{v^8} - \theta _c+\theta _p \right) -\frac{27}{4} \sin \left(\frac{3 \xi }{v^8} - 3 \theta _c-\theta _p \right) \\ -\frac{9}{32} \sin \left(\frac{3 \xi }{v^8} - 3 \theta _c+\theta _p \right) \Biggr\} - \frac{7}{384} \cos \left(\frac{\xi }{v^8} - \theta _c-\theta _p \right) -\frac{1}{288} \cos \left(\frac{\xi }{v^8} - \theta _c+\theta _p \right) \\ + \frac{9}{32} \cos \left(\frac{3 \xi }{v^8} - 3 \theta _c-\theta _p \right) + \frac{3}{256} \cos \left(\frac{3 \xi }{v^8} - 3 \theta _c+\theta _p \right) + \frac{625}{2304} \cos \left(\frac{5 \xi }{v^8} - 5 \theta _c-\theta _p \right) 
\end{multline}

\section{Jacobians}\label{app: jac_}
The Jacobian of the transformation from $(\mathcal{M},\, \eta,\, M_3,\, a) \to (\mathcal{M},\, \eta,\, z_{\rm L,0},\, \Omega_{\rm det})$ is given by
\begin{equation}
    \label{eq: jac_fin}
    \boldsymbol{J} = \frac{\partial (\mathcal{M}, \eta, z_{\rm L,0}, \Omega_{\rm det})}{\partial (\mathcal{M}, \eta, M_3, a)} = 
    \begin{pmatrix}
        1 & 0 & 0 & 0 \\
        0 & 1 & 0 & 0 \\
        - \frac{z_{\rm L,0} \eta^{-3/5}}{2(M_3 + M_{\rm S}) (1 + z_{\rm cos})} & \frac{3 z_{\rm L,0}}{10 \eta} \frac{M_{\rm S}}{M_3 + M_{\rm S}} & \frac{z_{\rm L,0}}{2} \frac{M_{\rm S}}{M_3 (M_3 + M_{\rm S})} & - \frac{z_{\rm L,0}}{2 a} \\
        \frac{\Omega_{\rm det} \eta^{-3/5}}{2(M_3 + M_{\rm S}) (1 + z_{\rm cos})} & - \frac{3 \Omega_{\rm det}}{10 \eta} \frac{M_{\rm S}}{M_3 + M_{\rm S}} & - \frac{\Omega_{\rm det}}{2} \frac{2 M_3 + 3 M_{\rm S}}{M_3 (M_3 + M_{\rm S})} & - \frac{3 \Omega_{\rm det}}{2 a}
    \end{pmatrix}
\end{equation}

For $M_3 \gg M_{\rm S}$, the Jacobian of the transformation from $(M_3,\, a) \to (z_{\rm L,0},\, \Omega_{\rm det})$ is given by
\begin{equation}
    \label{eq: jac_for_inf}
    \boldsymbol{J}_2 =  \frac{\partial (z_{\rm L,0},\, \Omega_{\rm det})}{\partial (M_3, a)} = \begin{pmatrix}
        0 & - z_{\rm L,0}^3 \\
        - \frac{\Omega_{\rm det}}{M_3} & - 3 z_{\rm L,0}^2 \Omega_{\rm det}  \\
    \end{pmatrix}
\end{equation}

\section{Additional Waveform Examples}\label{app: add_fig_wav}
\begin{figure*}[!ht]
    \centering
    \includegraphics[width=0.975\linewidth]{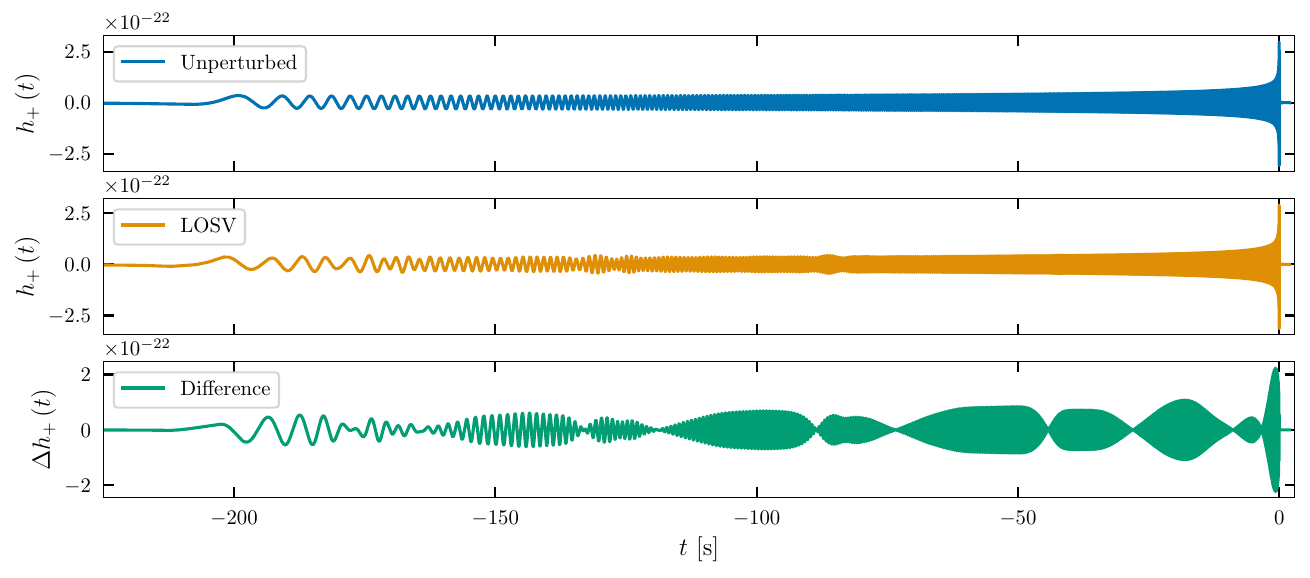}
    \caption{\textbf{Example Waveform}: The {\it upper panel} shows the time domain waveform of the non-spinning static BBH considered in Figure~\ref{fig: waveform_ex}, the {\it middle panel} shows the same when the outer orbit is eccentric having eccentricity 0.5 and semi-major axis $2.25 \times 10^3 \, R_{\rm s}$, and the {\it bottom panel} shows the difference between the two waveforms.}
    \label{fig: waveform_ex_eo}
\end{figure*}

\begin{figure*}[!ht]
    \centering
    \includegraphics[width=0.975\linewidth]{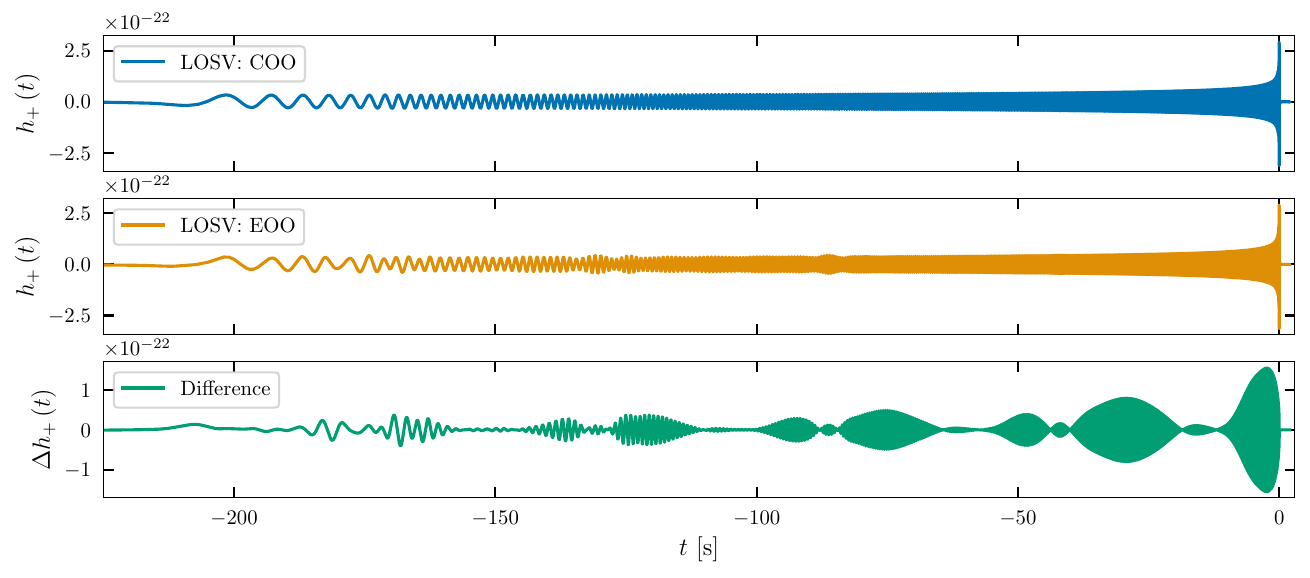}
    \caption{\textbf{Example Waveform}: The {\it upper panel} shows the time domain waveform of the perturbed BBH considered in Figure~\ref{fig: waveform_ex} (COO), the {\it middle panel} shows the same perturbed BBH considered in Figure~\ref{fig: waveform_ex_eo} (EOO), and the {\it bottom panel} shows the difference between the two waveforms.}
    \label{fig: waveform_ex_co_eo_diff}
\end{figure*}

Figure~\ref{fig: waveform_ex_eo} shows the time domain waveforms for the system considered in Figure~\ref{fig: waveform_ex} for an eccentric outer orbit of eccentricity 0.5 with $a$ being the semi-major axis and $\theta_{\rm c} = \vartheta_{\rm p} = 0.1$ radians. We find the {\it match} between the unperturbed and perturbed waveforms in A+, in this case, to be 0.608. Figure~\ref{fig: waveform_ex_co_eo_diff} shows the comparison between the two perturbed waveforms in the cases of circular and eccentric outer orbits. We find the {\it match} between the two waveforms in A+ to be 0.908.

\begin{figure}
    \centering
    \includegraphics[width=0.975\linewidth]{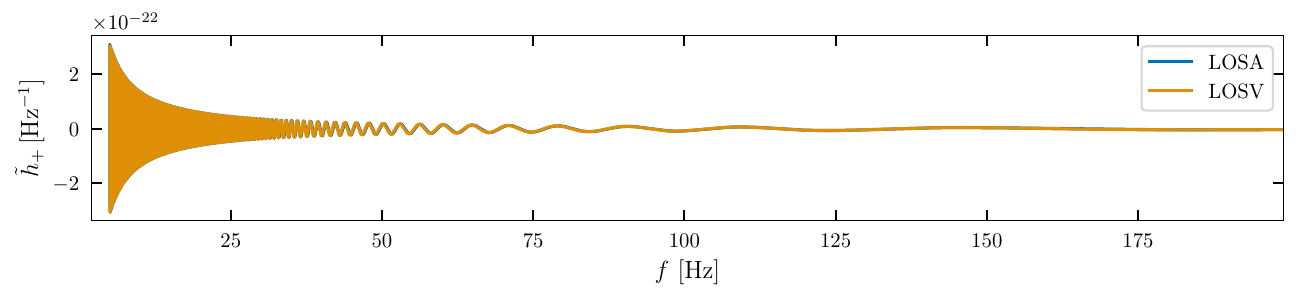}
    \includegraphics[width=0.975\linewidth]{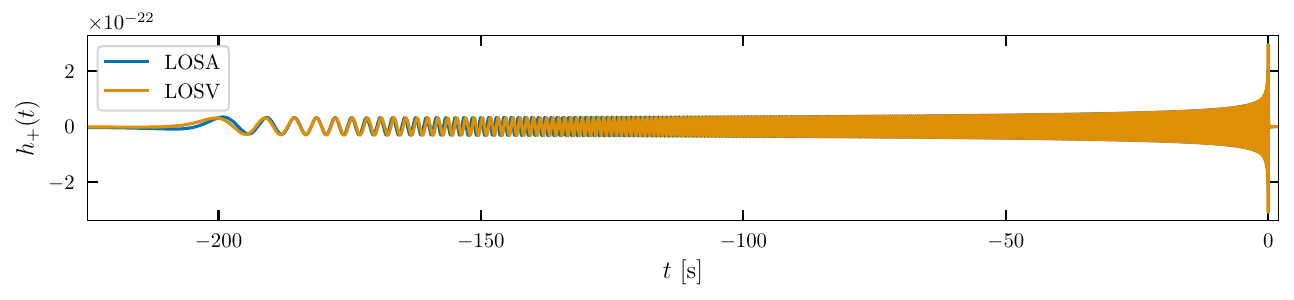}
    \caption{The {\it upper panel} shows the frequency domain waveforms of the A+: BBH system considered in Table~\ref{tab: tab_sys} in presence of a $8 \, M_{\odot}$ BH in the vicinity at $a = 3 \times 10^4 \, R_{\rm s}$ in COO scenario after incorporating only LOSA corrections (blue) in the waveform and full LOSV corrections (orange), while the {\it middle panel} shows that of the same in time domain.}
    \label{fig: waveform_ex_co_3e4}
\end{figure}

\begin{figure}
    \centering
    \includegraphics[width=0.75\linewidth]{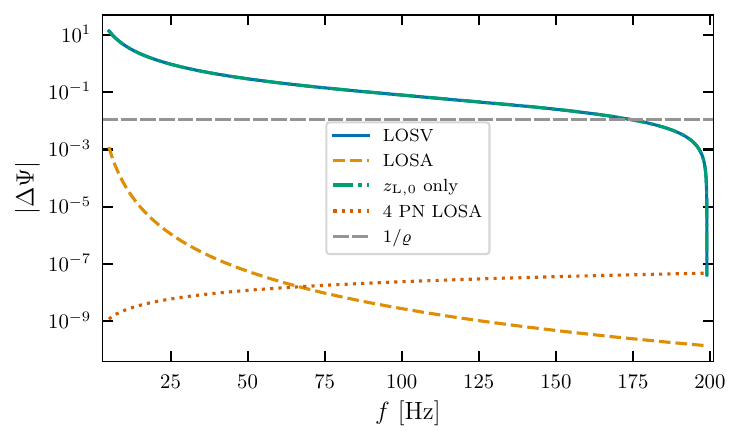}
    \caption{A comparison of the phase corrections due to LOSV, LOSA, and other terms appearing in the expansion of Equation~\eqref{eq: ph_cor_losv_circ} in the limit $\xi / v^8 \ll 1$ corresponding to the scenario considered in Figure~\ref{fig: waveform_ex_co_3e4}.}
    \label{fig: waveform_ex_co_3e4_dpsi}
\end{figure}

Figure~\ref{fig: waveform_ex_co_3e4} shows a comparison of the frequency ({\it top panel}) and time ({\it bottom panel}) domain waveforms for A+: BBH system in presence of a $8 \, M_{\odot}$ BH in the vicinity at $a = 3 \times 10^4 \, R_{\rm s}$ in COO scenario after incorporating only LOSA corrections (blue) and LOSV corrections (orange) --- this configuration leads to $z_{\rm L,0} = 2.2 \times 10^{-3}, \, \Omega_{\rm det} = 2.9 \times 10^{-3} \, {\rm Hz}, \, T_{\rm out} = 2148.7 \, {\rm s},$ and LOSA $\Gamma_1 = -1.9 \times 10^{-9} \, {\rm s}^{-1}$. Notice that even though both waveforms appear to be similar in the frequency domain, they differ in the time domain. To understand this, we Taylor Expand equation~\eqref{eq: ph_cor_losv_circ} in the limit $\xi / v^8 \ll 1$ to obtain
\begin{equation}
    \label{eq: ph_cor_losv_circ_Texp}
    \Delta \Psi_{\rm LC} = - \frac{5}{128 \eta v^5} z_{\rm L,0} \cos \theta _{\rm c} \left( 1-  \frac{v^8 }{v_{\rm lso}^8} \right) + \frac{25 v^3 }{65536 \eta ^2 v_{\rm lso}^{16}} \frac{G M}{c^3}  z_{\rm L,0} \Omega_{\rm det} \sin \theta _{\rm c} - \frac{25}{65536 \eta ^2 v^{13}} \frac{G M}{c^3}  z_{\rm L,0} \Omega_{\rm det} \sin \theta _{\rm c} + \mathcal{O} \left( \frac{\xi}{v^8} \right).
\end{equation}
The first term of this equation is $\Omega_{\rm det}$ independent/$z_{\rm L,0}$ only term, the second term is a 4 PN LOSA term, while the third term is the $-4$ PN LOSA term, where $- z_{\rm L,0} \Omega_{\rm det} \sin \theta _{\rm c}$ is the LOSA.
Figure~\ref{fig: waveform_ex_co_3e4_dpsi} shows a comparison of phase corrections due to LOSV (solid line), LOSA (dashed orange line), 4 PN LOSA term (dotted line), and $z_{\rm L,0}$ only term (dash-dotted line). The dashed horizontal line in the same figure shows the minimum phase shift that one can measure for this signal, which is provided by 1/SNR. Notice that the phase shift due to LOSV is measurable while the same due to LOSA is not. Given that we are in $\xi / v^8 \ll 1$ regime, the main contribution to LOSV corrections must be coming from $z_{\rm L,0}$ only terms if LOSA corrections are not measurable, which is what we see in Figure~\ref{fig: waveform_ex_co_3e4_dpsi} ($z_{\rm L,0}$ only correction sitting on top of the full LOSV correction). However, this would not give us any additional information because when $\Omega_{\rm det}$ or, equivalently, LOSA (to the lowest order) is not measurable, even a periodic LOSV will lead to a constant Doppler shift, which will not be measurable due to mass-redshift degeneracy. It can be seen by taking the $\Omega_{\rm det} \to 0$, i.e., $\xi \to 0$ limit of equation~\eqref{eq: neg_coal_t_circ}:
\begin{equation}
    \label{eq: neg_coal_time_lim_om0}
    \lim_{\xi \to 0} \, (t - t_{\rm c})_{\rm LC} = -\frac{5}{256 \eta  v^8} \frac{G M}{c^3} \left( 1 - \frac{5}{3} z_{\rm L,0} \cos \theta_{\rm c} \right) + \frac{5}{256 \eta  v_{\rm lso}^8} \frac{GM}{c^3} z_{\rm L,0} \cos \theta_{\rm c}
\end{equation}
that the term in the brackets can be absorbed in the redefinition of $M$, while the second term, which is a constant, can be absorbed in the redefinition of $t_{\rm c}$. As a result, the measurability of phase correction due to LOSV in this case is a consequence of $M$ being measurable with a precision better than the change in $M$ due to LOSV, which is $z_{\rm L,0} M \cos \theta_{\rm c}$.

\section{Additional Figures}\label{app: add_fig}
\begin{figure}
    \centering
    \includegraphics[width=0.75\linewidth]{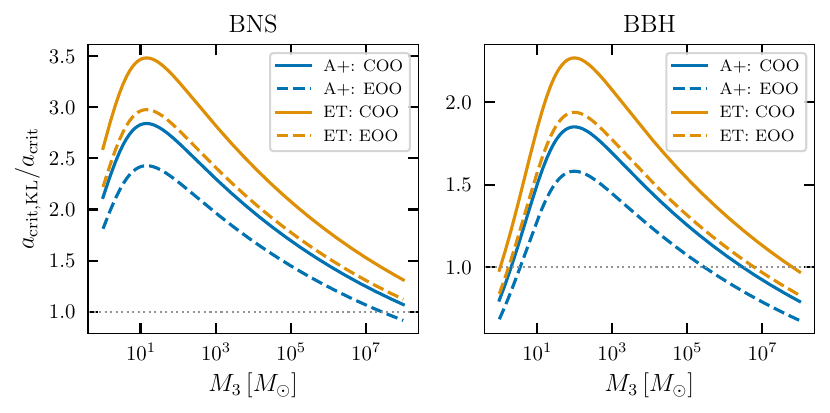}
    \caption{The variation of $a_{\rm crit,KL} / a_{\rm crit}$, equation~\eqref{eq: acrit_kl_acrit_comp}, with the mass of the tertiary in the vicinity of the BNS and BBH systems considered in A+ and ET (see Table~\ref{tab: tab_sys}).}
    \label{fig: acrit_comp}
\end{figure}

Figure~\ref{fig: acrit_comp} shows a comparison of $a_{\rm crit,KL}$ and $a_{\rm crit}$ given by equations~\eqref{eq: acrit_kl_def} and~\eqref{eq: stab_rs}, respectively, for a range of tertiary masses in the vicinity of the A+BNS, A+: BBH, ET: BNS, and ET: BBH systems (see Table~\ref{tab: tab_sys}). Notice that for the tertiary in the vicinity of the BNS considered in this work, $a_{\rm crit,KL}$ is always greater than $a_{\rm crit}$ except for $M_3 > 10^7 \, M_{\odot}$ in the EOO scenario, where the Doppler modulations are unmeasurable. For the tertiary in the vicinity of the BBH, $a_{\rm crit,KL}$ is always greater than $a_{\rm crit}$, except for $M_3 \lesssim 4 \, M_{\odot}$ and  $M_3 \gtrsim 10^5 \, M_{\odot}$ in A+; and for $M_3 \lesssim 2 \, M_{\odot}$ and  $M_3 \gtrsim 10^7 \, M_{\odot}$ in ET.

\begin{figure}
    \centering
    \includegraphics[width=0.495\linewidth]{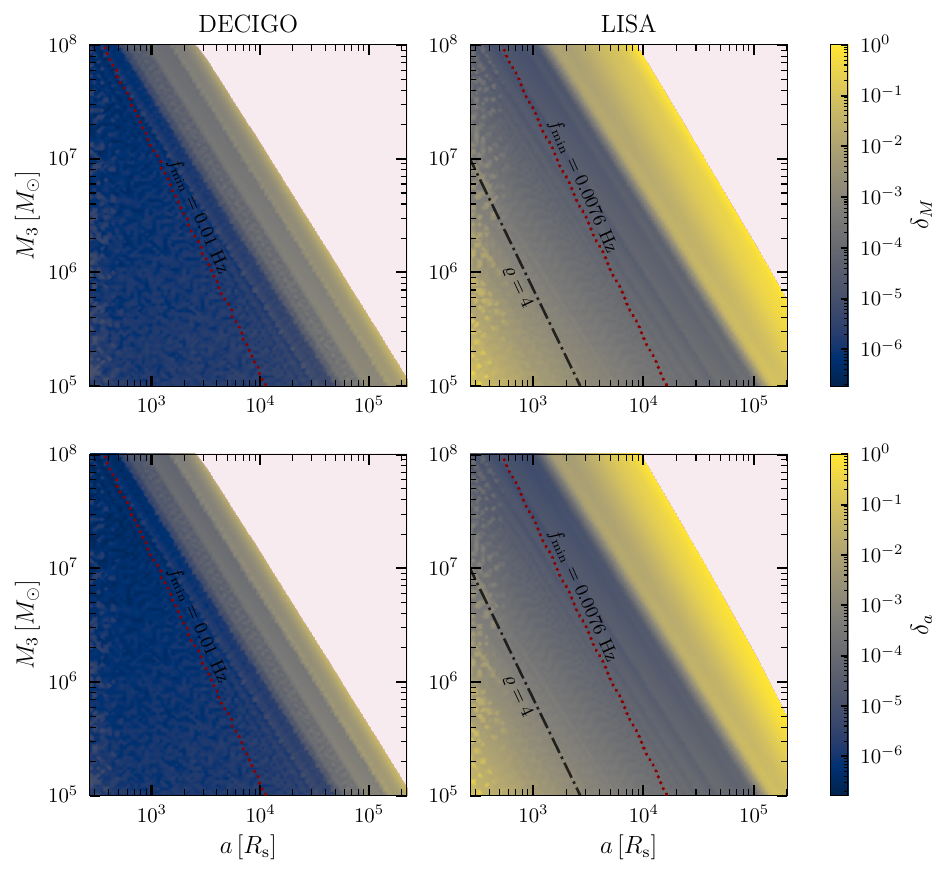}
    \includegraphics[width=0.495\linewidth]{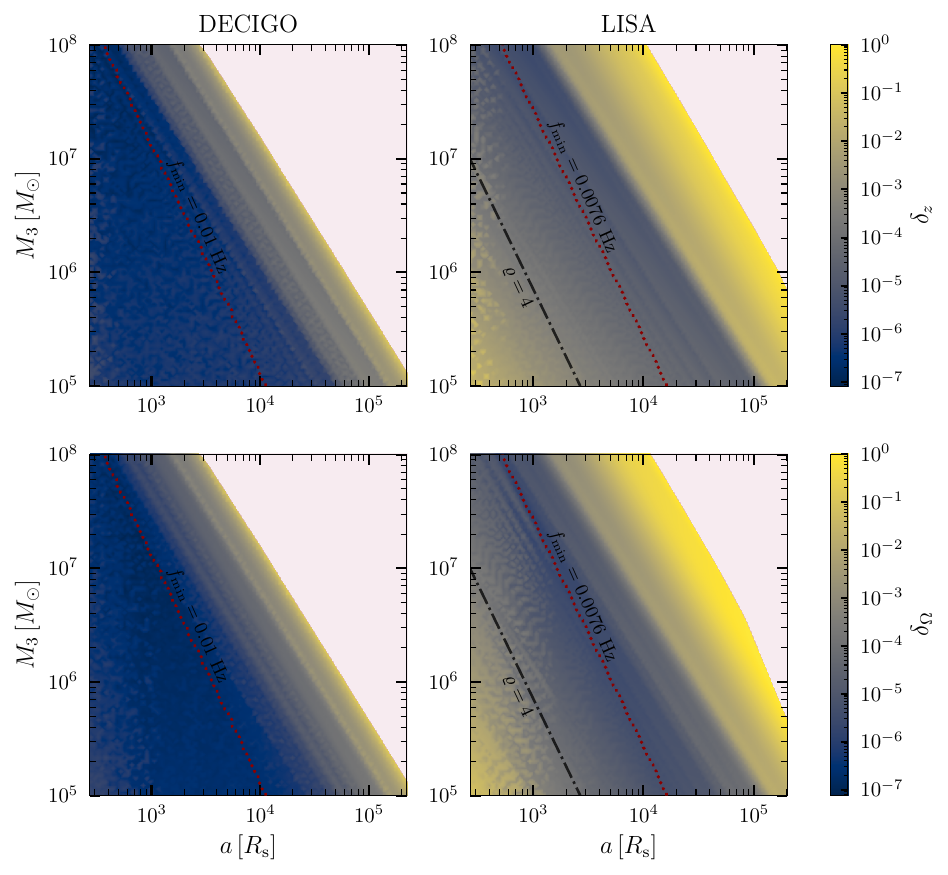}
    \caption{\textbf{SMBH:} The {\it left two panels} show the relative errors in the measurement of $M_3$ ({\it top panels}) and $a$ ({\it bottom panels}) over a grid of $M_3$ and $a$ for the DECIGO: BBH and LISA: BBH cases corresponding to top panels of Figures 3 and 4 of~\cite{Tiwari:2024pvb} in EOO scenario and the {\it right two panels} show the corresponding relative errors in the measurement of $z_{\rm L,0}$ ({\it top panels}) and $\Omega_{\rm det}$ ({\it bottom panels}). The patches on the upper right and the dotted lines have the same meaning as in Figure~\ref{fig: Ap_ET_BNS_BBH_SBH_IMBH_CO}, while the dashed-dotted lines have the same meaning as in Figure~\ref{fig: Ap_ET_DECIGO_LISA_CO_ME}.}
    \label{fig: DECIGO_LISA_SMBH_EO}
\end{figure}

Figure~\ref{fig: DECIGO_LISA_SMBH_EO} shows the relative error in $M_3$ and $a$ ({\it left two panels}) for DECIGO: BBH and LISA: BBH systems considered in Table~\ref{tab: tab_sys} in EOOs of eccentricity 0.5 with $\vartheta_{\rm p} = 0$ and varying $\theta_{\rm c}$ over the $M_3 - a$ (SMBH-like tertiary masses) grid to mimic the situation $a/r = 1.25$ considered in Figures 3 and 4 of~\cite{Tiwari:2024pvb}, $r$ being the location of the CBC in the outer orbit at coalescence. We find that a $10^5 \, M_{\odot}$ SMBH in the vicinity of a $100-100 \, M_{\odot}$ BBH at 1 Gpc can be detected in an EOO of eccentricity 0.5 and a semi-major axis greater than $\sim 2 \times 10^5 \, R_{\rm s} \, (394.83 \, \rm AU)$ with DECIGO and LISA both, while a $10^8 \, M_{\odot}$ SMBH in the same configuration can be detected up to $\sim 3 \times 10^3 \, R_{\rm s} \, (0.029 \, \rm pc)$ and $\sim 10^4 \, R_{\rm s} \, (0.096 \, \rm pc)$ with DECIGO and LISA, respectively. Note that for both cases, as we enter in the deep $\Omega_{\rm det} t_{\rm obs} / 2 \pi \ll 1$ region such that only a few terms of the sine and cosine series expansion in Equation~\eqref{eq: ph_cor_losv_ecc} become dominant, the Fisher matrix inversion in terms of $z_{\rm L,0}$, $\ln \Omega_{\rm det}$, and $e_{\rm out}$ becomes inefficient because the matrix becomes roughly singular and one would be required to parametrize the kinematics of the CBC's CoM in terms of LOSA and its higher order time derivatives. 

\begin{figure*}
    \centering
    \includegraphics[width=0.495\linewidth]{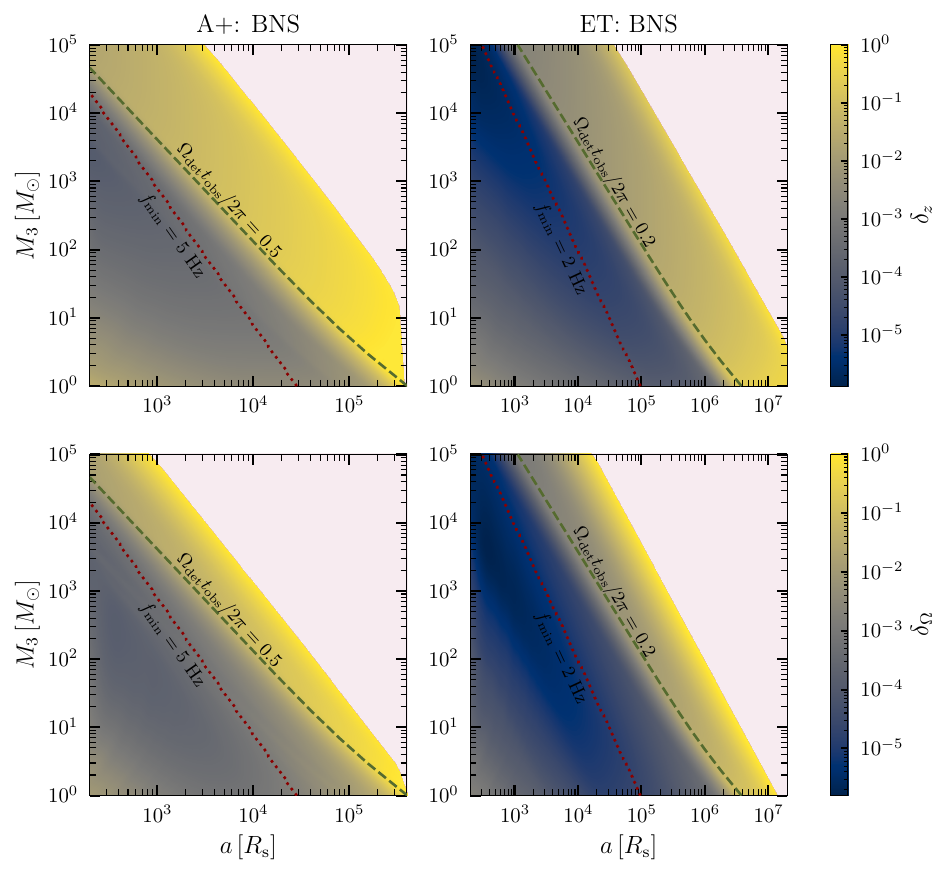}
    \includegraphics[width=0.495\linewidth]{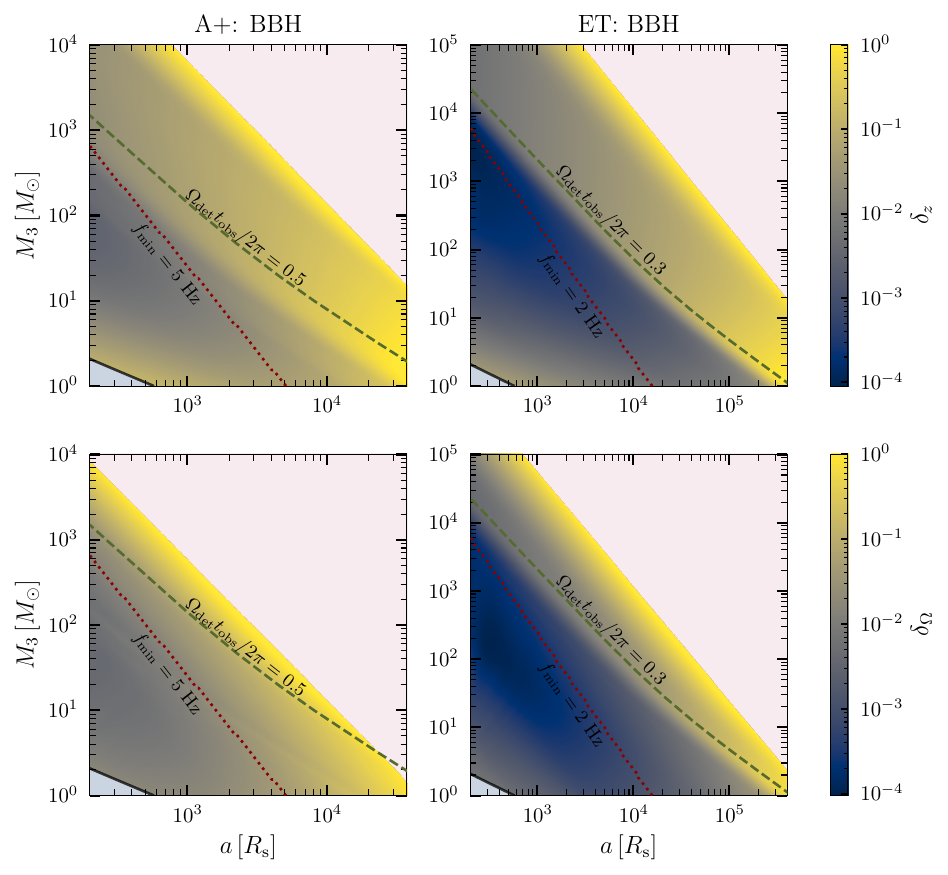}
    \caption{\textbf{SBH-IMBH:} The {\it left two panels} show the relative errors in the measurement of $z_{\rm L,0}$ ({\it top panels}) and $\Omega_{\rm det}$ ({\it bottom panels}) over a grid of $M_3$ and $a$ for the A+: BNS and ET: BNS cases corresponding to ~\ref{fig: Ap_ET_BNS_BBH_SBH_IMBH_CO} in COO scenario, while the {\it right two panels} show the same for A+: BBH and ET: BBH cases. The patches on the upper right and bottom left, and the dashed and dotted lines have the same meaning as in Figure~\ref{fig: Ap_ET_BNS_BBH_SBH_IMBH_CO}.}
    \label{fig: Ap_ET_BNS_BBH_SBH_IMBH_zo_CO}
\end{figure*}

Figure~\ref{fig: Ap_ET_BNS_BBH_SBH_IMBH_zo_CO} shows the relative errors in the measurement of $z_{\rm L,0}$ and $\Omega_{\rm det}$ over the $M_3 - a$ (SBH- and IMBH-like tertiary masses) grid for A+: BNS, ET: BNS, A+: BBH, and ET: BBH cases corresponding to Figure~\ref{fig: Ap_ET_BNS_BBH_SBH_IMBH_CO} in COO scenario. 

\begin{figure}[ht!]
    \centering
    \includegraphics[width=0.5\linewidth]{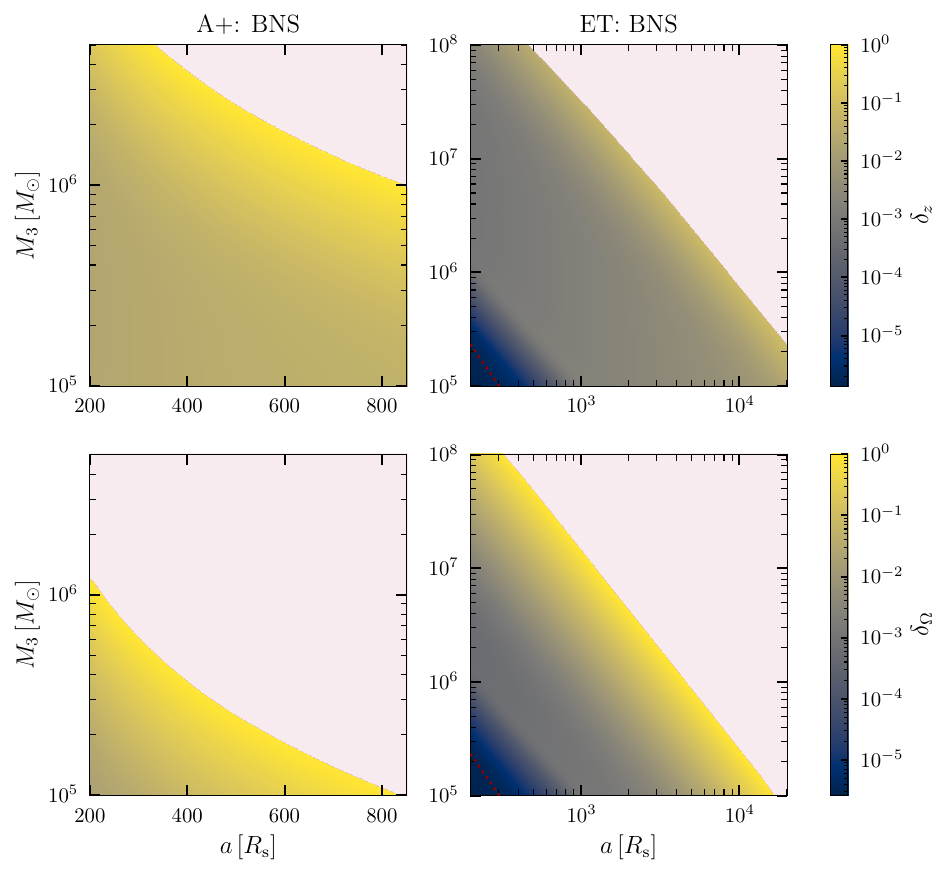}
    \caption{\textbf{SMBH:} The {\it left} and {\it right panels} show the relative errors in the measurement of $z_{\rm L,0}$ ({\it top panels}) and $\Omega_{\rm det}$ ({\it bottom panels}) over a grid of $M_3$ and $a$ for the A+: BNS and ET: BNS cases corresponding to Figure~\ref{fig: Ap_ET_bns_SMBH} in COO scenario. The patches on the upper right and the dotted lines have the same meaning as in Figure~\ref{fig: Ap_ET_BNS_BBH_SBH_IMBH_CO}.}
    \label{fig: Ap_ET_bns_SMBH_zo_CO}
\end{figure}

Figure~\ref{fig: Ap_ET_bns_SMBH_zo_CO} shows the relative errors in the measurement of $z_{\rm L,0}$ and $\Omega_{\rm det}$ over the $M_3 - a$ (SMBH-like tertiary masses) grid for A+: BNS and ET: BNS cases corresponding to Figure~\ref{fig: Ap_ET_bns_SMBH} in the COO scenario.

\begin{figure*}
    \centering
    \includegraphics[width=0.495\linewidth]{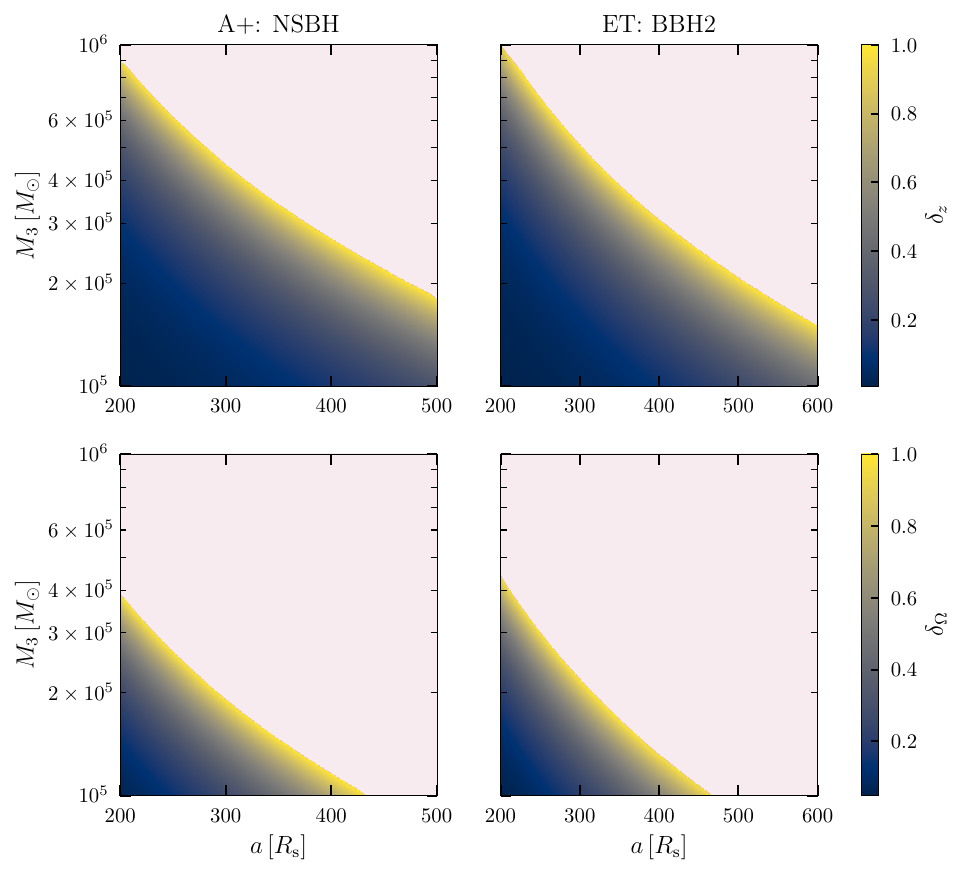}
    \includegraphics[width=0.495\linewidth]{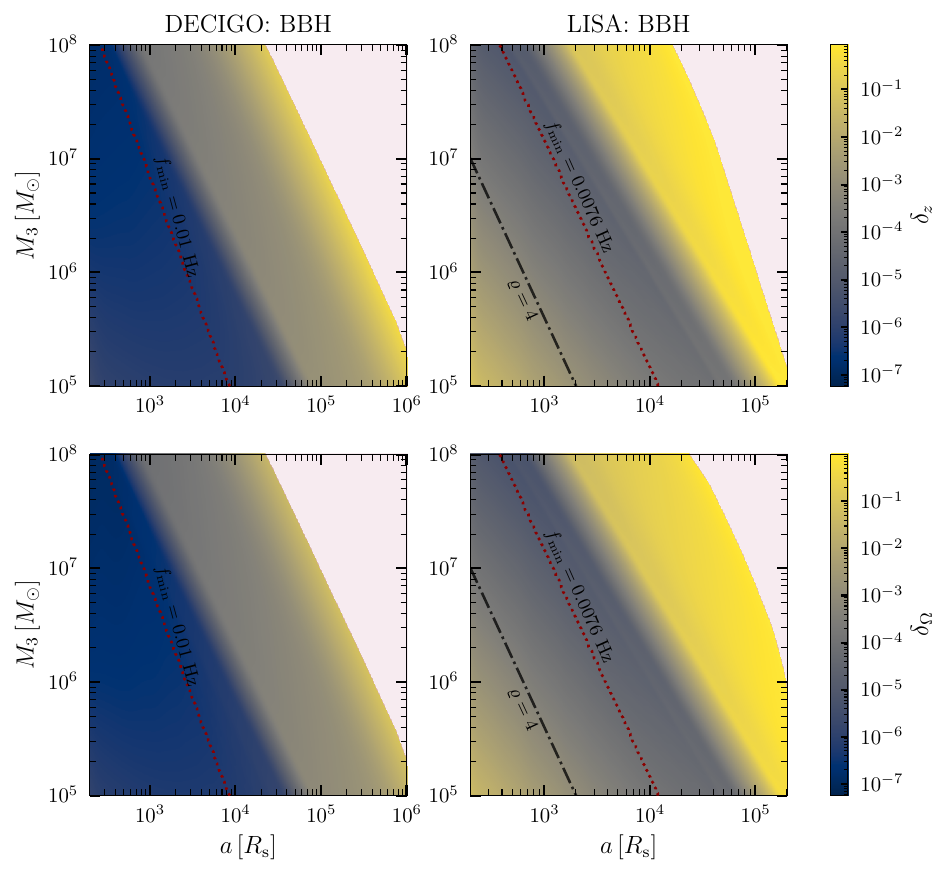}
    \caption{\textbf{SMBH:} The {\it left two panels} show the relative errors in the measurement of $z_{\rm L,0}$ ({\it top panels}) and $\Omega_{\rm det}$ ({\it bottom panels}) over a grid of $M_3$ and $a$ for the A+: NSBH and ET: BBH2 cases corresponding to ~\ref{fig: Ap_ET_DECIGO_LISA_CO_ME} in COO scenario, while the {\it right two panels} show the same for DECIGO: BBH and LISA: BBH cases. The patches on the upper right and the dotted lines have the same meaning as in Figure~\ref{fig: Ap_ET_BNS_BBH_SBH_IMBH_CO}, while the dashed-dotted lines have the same meaning as in Figure~\ref{fig: Ap_ET_DECIGO_LISA_CO_ME}.}
    \label{fig: Ap_ET_DECIGO_LISA_CO_co_ME}
\end{figure*}

Figure~\ref{fig: Ap_ET_DECIGO_LISA_CO_co_ME} shows the relative errors in the measurement of $z_{\rm L,0}$ and $\Omega_{\rm det}$ over the $M_3 - a$ grid for A+: NSBH, ET: BBH2, DECIGO: BBH, and LISA: BBH cases corresponding to Figure~\ref{fig: Ap_ET_DECIGO_LISA_CO_ME} in the COO scenario.

\begin{figure*}
    \centering
    \includegraphics[width=0.495\linewidth]{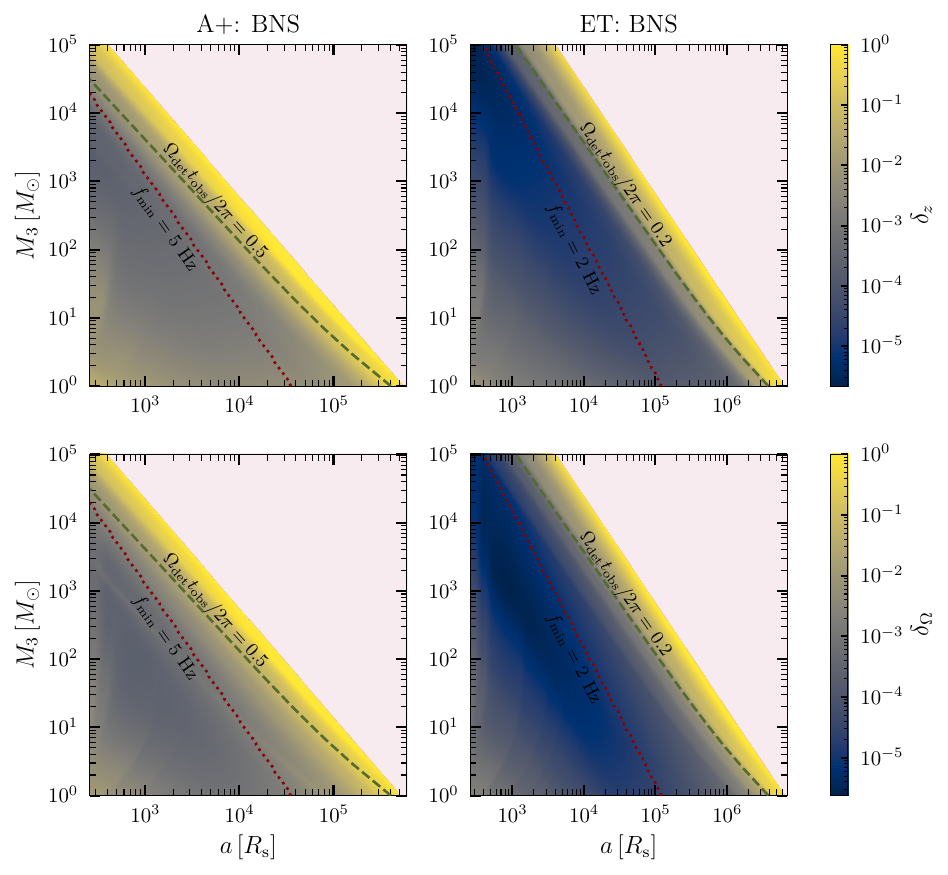}
    \includegraphics[width=0.495\linewidth]{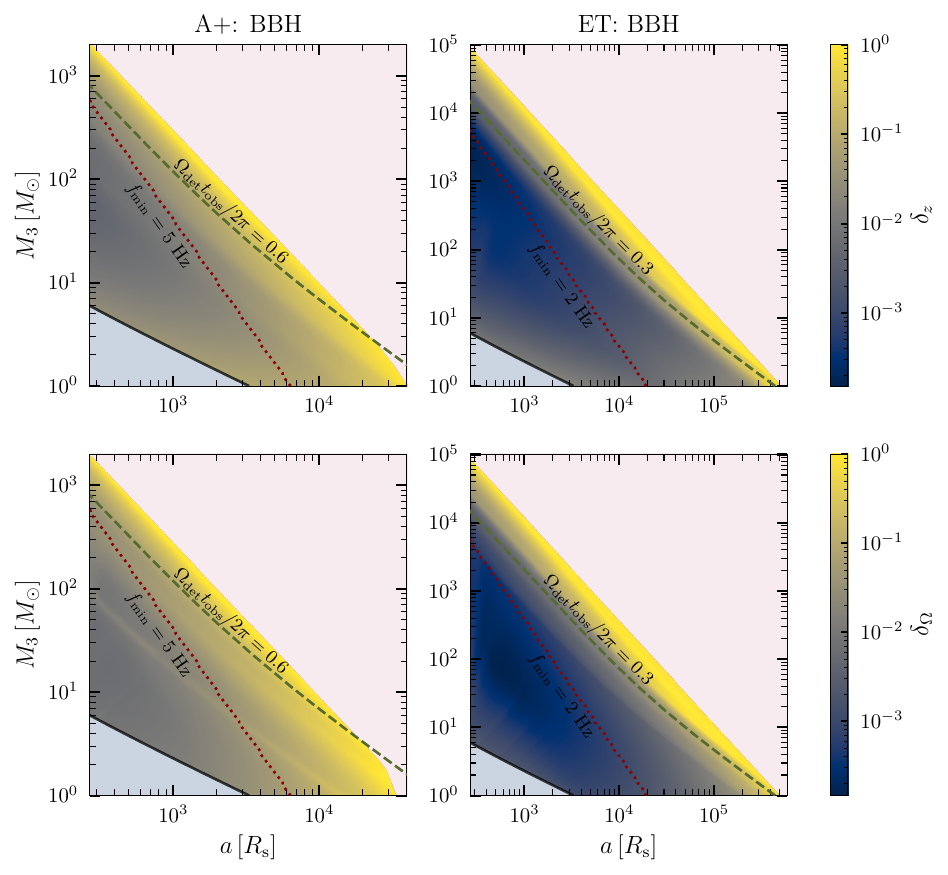}
    \caption{\textbf{SBH-IMBH:} The {\it left two panels} show the relative errors in the measurement of $z_{\rm L,0}$ ({\it top panels}) and $\Omega_{\rm det}$ ({\it bottom panels}) over a grid of $M_3$ and $a$ for the A+: BNS and ET: BNS cases corresponding to Figure~\ref{fig: Ap_ET_BNS_BBH_SBH_IMBH_EO} in EOO scenario, while the {\it right two panels} show the same for A+: BBH and ET: BBH cases. The patches on the upper right and the dotted lines have the same meaning as in Figure~\ref{fig: Ap_ET_BNS_BBH_SBH_IMBH_CO}, while the dashed-dotted lines have the same meaning as in Figure~\ref{fig: Ap_ET_DECIGO_LISA_CO_ME}.}
    \label{fig: Ap_ET_BNS_BBH_SBH_IMBH_zo_EO}
\end{figure*}

Figure~\ref{fig: Ap_ET_BNS_BBH_SBH_IMBH_zo_EO} shows the relative errors in the measurement of $z_{\rm L,0}$ and $\Omega_{\rm det}$ over the $M_3 - a$ (SBH- and IMBH-like tertiary masses) grid for A+: BNS, ET: BNS, A+: BBH, and ET: BBH cases corresponding to Figure~\ref{fig: Ap_ET_BNS_BBH_SBH_IMBH_EO} in the EOO scenario. 

\begin{figure*}
    \centering
    \includegraphics[width=0.8\linewidth]{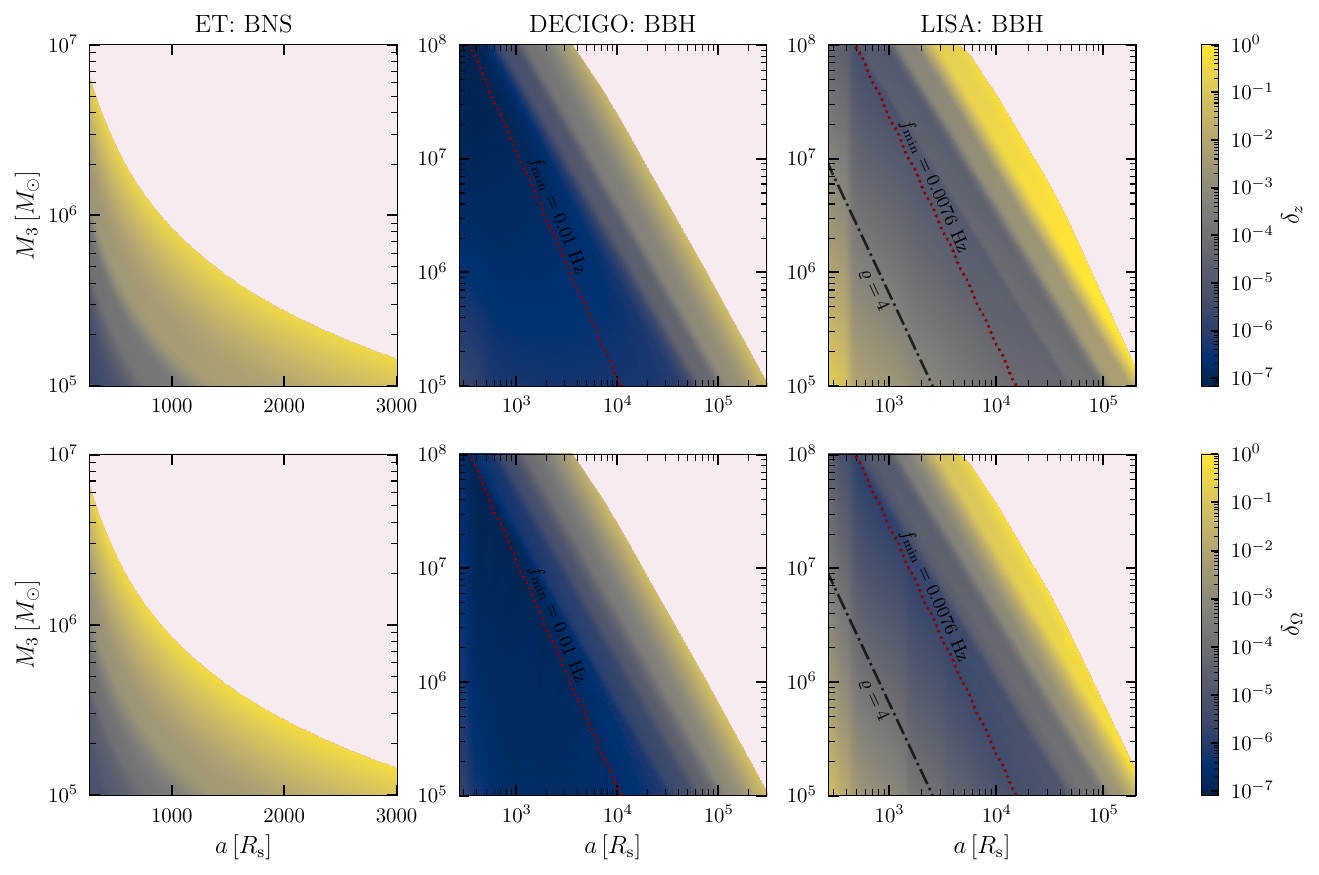}
    \caption{\textbf{SMBH:} The {\it left}, {\it middle}, and {\it right panels} show the relative errors in the measurement of $z_{\rm L,0}$ ({\it top panels}) and $\Omega_{\rm det}$ ({\it bottom panels}) over a grid of $M_3$ and $a$ for the ET: BNS, DECIGO: BBH, and LISA: BBH cases, respectively, corresponding to ~\ref{fig: ET_DECIGO_LISA_SMBH} in EOO scenario. The patches on the upper right have the same meaning as in Figure~\ref{fig: Ap_ET_BNS_BBH_SBH_IMBH_CO}.}
    \label{fig: ET_DECIGO_LISA_SMBH_zo_EO}
\end{figure*}

Figure~\ref{fig: ET_DECIGO_LISA_SMBH_zo_EO} shows the relative errors in the measurement of $z_{\rm L,0}$ and $\Omega_{\rm det}$ over the $M_3 - a$ (SMBH-like tertiary masses) grid for ET: BNS, DECIGO: BBH, and LISA: BBH cases corresponding to Figure~\ref{fig: ET_DECIGO_LISA_SMBH} in the EOO scenario.

\end{document}